\documentclass[twocolumn,trackchanges]{aastex701}
\usepackage{amsmath}
\usepackage{amssymb}

\shorttitle{Dynamical Forecasts for the Hierarchical Triple PSR~J0435$+$3233 System}
\shortauthors{Gopakumar et al.}

\defcitealias{YangHT}{Y26}
\defcitealias{JR97}{JR97}
\defcitealias{Gaia2016}{Gaia Collaboration et al. 2016}
\defcitealias{GaiaDR32022}{Gaia Collaboration et al. 2023}

\begin{document}

\title{
A Unified Multi-Band Secular Framework for the Hierarchical Triple PSR~J0435$+$3233 System
}

\author[orcid=0000-0003-4274-4369]{Achamveedu Gopakumar}
\affiliation{
Department of Astronomy and Astrophysics, Tata Institute of Fundamental Research, Mumbai 400005, Maharashtra, India
}
\email[show]{gopu.tifr@gmail.com}

\author[orcid=0009-0001-2049-3039]{Vidit Singh}
\affiliation{
Department of Astronomy and Astrophysics, Tata Institute of Fundamental Research, Mumbai 400005, Maharashtra, India
}
\email[show]{vidittt.singh@gmail.com}

\author[0000-0002-2820-0931]{Abhimanyu Susobhanan}
\affiliation{
School of Physics, Indian Institute of Science Education and Research Thiruvananthapuram, Maruthamala P.O., Thiruvananthapuram 695551, Kerala, India
}
\email{abhisrkckl@gmail.com}

\author[0000-0001-8580-8874]{Mauri J. Valtonen}
\affiliation{
FINCA, University of Turku, FI-20014 Turku, Finland
}
\affiliation{
Tuorla Observatory, Department of Physics and Astronomy, University of Turku, FI-20014 Turku, Finland
}
\email{mvaltonen2001@yahoo.com}

\author[0000-0001-8572-8241]{Sayantan Bhattacharya}
\affiliation{
Department of Astronomy and Astrophysics, Tata Institute of Fundamental Research, Mumbai 400005, Maharashtra, India
}
\email{sayantan.bhattacharya@tifr.res.in}  

\author[0000-0003-3010-7661]{Di Li}
\affiliation{
New Cornerstone Science Laboratory, Department of Astronomy, Tsinghua University, Beijing 100084, China
}
\affiliation{
National Astronomical Observatories, Chinese Academy of Sciences, Beijing 100101, China
}
\email{dili@tsinghua.edu.cn} 

\author[0000-0002-8609-3599]{Hong-Liang Yan}
\affiliation{
National Astronomical Observatories, Chinese Academy of Sciences, Beijing 100101, China
}
\email{hlyan@nao.cas.cn}

\author[0000-0002-5381-6498]{Jianping Yuan}
\affiliation{Xinjiang Astronomical Observatory, Chinese Academy of Sciences, Urumqi, Xinjiang 830011, China}
\email{yuanjp@xao.ac.cn}

\author[0000-0002-9786-8548]{Na Wang}
\affiliation{Xinjiang Astronomical Observatory, Chinese Academy of Sciences, Urumqi, Xinjiang 830011, China}
\email{na.wang@xao.ac.cn}

\author[0000-0002-3386-7159]{Pei Wang}
\affiliation{
National Astronomical Observatories, Chinese Academy of Sciences, Beijing 100101, China
}
\email{wangpei@nao.cas.cn} 

\author[0000-0002-1495-4011]{Qingdong Wu}
\affiliation{
School of Physics and Electronic Science, Zunyi Normal University, Zunyi 563006, China
}
\email{wuqingdong@zync.edu.cn}

\author[0009-0007-8062-1454]{De Zhao}
\affiliation{Xinjiang Astronomical Observatory, Chinese Academy of Sciences, Urumqi, Xinjiang 830011, China}
\email{zhaode@xao.ac.cn}

\begin{abstract}

The millisecond pulsar PSR~J0435$+$3233, discovered by FAST, exhibits an unusually large apparent spin-down rate, potentially reflecting line-of-sight acceleration from a distant companion. Recent investigations reveal a hierarchical triple consisting of an inner pulsar--helium white dwarf binary with an orbital period of $\sim 8$ days, orbited by a G-type subgiant tertiary on an eccentric $\sim 74$ yr orbit. The current timing description admits two distinct geometrical solutions. We develop a singly averaged (SA) formulation that averages over the inner orbit while retaining the instantaneous phase of the eccentric outer orbit, thereby capturing the time-dependent quadrupolar perturbations induced by the tertiary. We show how this physical three-body description can be incorporated naturally into pulsar-timing analyses and use it to forward-propagate the two currently viable configurations, Tv1 and Tv2, to forecast when their timing signatures become distinguishable with FAST. The SA evolution predicts that the two configurations will become distinguishable in the coming years, with the effective projected semi-major axis, $x_{1,\rm eff}$, providing the earliest discriminating observable. We further extend the framework to predict the tertiary's spectroscopic radial-velocity and astrometric evolution, linking FAST timing with LAMOST time-domain spectroscopy and \textit{Gaia} astrometry. This multi-band framework has the potential to establish PSR~J0435$+$3233 as an exceptionally well-constrained laboratory for three-body dynamics, compact-binary and stellar evolution, and gravitational physics.

\end{abstract}


\section{Introduction}
\label{sec:intro}

Hierarchical triple (HT) systems containing millisecond pulsars (MSPs) are extraordinarily rare astrophysical configurations that serve as high-precision laboratories for testing stellar evolution, binary recycling mechanisms, and alternative theories of gravity \citep{Ransom2014, Shao2016,Archibald2018, Iorio2024}. To date, only two such systems have been extensively characterized in the Galactic field or clusters. The first, PSR~B1620$-$26, resides in a globular cluster and is accompanied by a white dwarf and an ultra-light, planetary-mass tertiary companion on a wide outer orbit \citep{Thorsett1999}. The second, the famous double-white-dwarf system PSR~J0337$+$1715, contains two white dwarfs in  nearly coplanar, and circular orbits \citep{Ransom2014}. Because of the extreme differences in gravitational binding energy between its compact components, PSR~J0337$+$1715 has stood as the gold standard for testing the Strong Equivalence Principle (SEP) in the strong-field regime \citep{Archibald2018,Voisin2020, Voisin2025}. Therefore, discovering new, highly asymmetric hierarchical configurations remains essential to probing the full spectrum of three-body dynamics and non-coplanar stellar evolution.

This observational landscape has been fundamentally expanded by the discovery of the millisecond radio pulsar PSR~J0435$+$3233 ($P \simeq  3.20\,\rm ms$) by \citet{Wu2026} using the Five-hundred-meter Aperture Spherical radio Telescope \citep[FAST;][]{FASTref} during the Commensal Radio Astronomy FAST Survey \citep[CRAFTS;][]{Crafts2018}. Long-term timing by \citet{Wu2026} revealed that the pulsar moves in a circular 8-day inner orbit with a likely helium white dwarf (He~WD) companion of mass $\sim0.3\,M_\odot$, but displays an exceptionally large and rapidly varying orbital period derivative. Subsequent independent investigations by \citet{Freire2026} and \citet{YangHT} (hereafter \citetalias{YangHT}) successfully resolved this anomalous motion.

Combining a reanalysis of the FAST archive with $\sim$16.7 years of $\gamma$-ray pulsations from the \textit{Fermi} Large Area Telescope \citep[LAT;][]{FermiLAT, Fermi4FGL2, Fermi4FGL} and astrometric information from \textit{Gaia} (\citetalias{Gaia2016}), \citetalias{YangHT} established the triple nature of PSR~J0435$+$3233 and associated the bright optical/infrared source Gaia~DR3~171833395176138240 with its tertiary companion. The latter is a Sun-like G-type subgiant with an inferred mass of $M_\star \simeq0.98$--$1.12\,M_\odot$, orbiting the inner pulsar--white dwarf binary on a wide ($P_2\simeq74\,{\rm yr}$) and highly eccentric ($e_2\simeq0.5983$) outer orbit. This effort established PSR~J0435$+$3233 as only the third known MSP residing in a hierarchical triple system, and only the second such system identified in the Galactic field (\citetalias{YangHT}).

To map the three-dimensional architecture of the system, \citetalias{YangHT} combined high-precision FAST radio pulse times of arrival (ToAs) with $\sim$16.7 years of \textit{Fermi}-LAT gamma-ray pulsation timing. These timing data sets were further complemented by \textit{Gaia} DR3 (\citetalias{GaiaDR32022}) astrometric, distance, and mass constraints on the subgiant counterpart. By modeling the tertiary-induced non-Keplerian variations of the inner orbital parameters using these datasets, they constrained the outer orbital parameters. Specifically, \citetalias{YangHT}'s joint Bayesian analysis using nested-sampling  yielded the following two degenerate three-dimensional geometries:
\begin{itemize}
    \item \textbf{Solution Tv1}: Characterized by a nearly perpendicular mutual orbital inclination of $i_{\rm X} \simeq 83.8^\circ$ between the inner and outer orbits, with component masses of $M_{\rm PSR} \approx 1.15\,M_\odot$, $M_{\rm WD} \approx 0.271\,M_\odot$, and a subgiant tertiary mass of $M_{\star} \approx 0.96\,M_\odot$.
    \item \textbf{Solution Tv2}: Characterized by a moderate mutual inclination of $i_{\rm X} \simeq 55.0^\circ$, with corresponding component masses of $M_{\rm PSR} \approx 1.29\,M_\odot$, $M_{\rm WD} \approx 0.296\,M_\odot$, and $M_{\star} \approx 1.12\,M_\odot$.
\end{itemize}

It is important to note that while both solutions describe the current baseline of FAST timing observations with similar timing residual statistics, they imply fundamentally different orbital configurations and mass hierarchies (\citetalias{YangHT}). Because the tertiary star is currently moving toward its next periastron passage, scheduled for mid-2036 as evident from Table~4 of \citetalias{YangHT}, the rapidly strengthening tidal perturbation exerted by the outer orbit on the inner binary, and the resulting evolution of the inner-orbit geometry, provide a natural opportunity to break this degeneracy.

Building directly upon this breakthrough by \citetalias{YangHT}, we employ a singly averaged (SA) formulation of HT dynamics in which the rapid motion of the inner binary is orbit-averaged, while the instantaneous position of the tertiary along its eccentric outer orbit is retained. This allows us to incorporate naturally the time-dependent quadrupolar interaction between the inner and outer orbits, influenced by \citet{Luo2016, Toonen2016,Hamers2020}. The SA formulation turned out to be particularly well suited for describing the observationally relevant dynamics of the HT associated with PSR~J0435$+$3233. Specifically, it predicts the continuous evolution of the inner orbital geometry, which can be mapped directly onto pulsar timing observables such as the projected semi-major axis and its temporal variation. The resulting dynamical timing model can therefore be incorporated naturally into the analysis of radio pulse ToAs measured using FAST. It replaces the phenomenological description of orbital evolution through higher-order time derivatives of the inner orbital elements, as pursued in \citetalias{YangHT}, with a physically motivated and self-consistent evolution governed by the underlying three-body dynamics.

Thereafter, we employ our SA framework to forward propagate the competing Tv1 and Tv2 solutions, construct their corresponding FAST timing forecasts, and determine when their predicted timing signatures should become statistically distinguishable with continued observations. Interestingly, our SA framework reveals that, although the Tv1 and Tv2 solutions remain nearly degenerate over the existing $\sim$4.6-yr FAST timing baseline, their predicted trajectories should become distinguishable well before the 2036 outer periastron. The separation appears first in $x_{1,\mathrm{eff}}$, which is particularly sensitive to the different inclination evolution of the two three-dimensional configurations, while differences in the eccentricity-vector components $\epsilon_1$ and $\epsilon_2$ become prominent subsequently. We forecast that extending FAST timing by only another 2 years,  will produce a separation of $\Delta x_{1,\mathrm{eff}}\sim 100\,\mu{\rm s}$ between the two predicted trajectories. This also explains why the configurations could not be distinguished with the existing data: over the current baseline, both remain locally well described by the phenomenological Taylor expansions employed in \citetalias{YangHT}, whereas their subsequent nonlinear evolution increasingly departs from such local descriptions as the tertiary advances toward periastron. Continued FAST timing should therefore distinguish the two physical configurations in the near future, without requiring observations extending to the 2036 periastron passage.

Additionally, we demonstrate how our SA framework provides a natural route toward a unified multi-band description of PSR~J0435$+$3233. The three-dimensional relative separation and velocity vectors of the outer orbit, $\mathbf{R}(t)$ and $\mathbf{V}(t)$, connect the same dynamical model directly to the spectroscopic radial-velocity and astrometric signatures of the optically identified tertiary, Gaia~DR3~171833395176138240. This opens the possibility of a joint Bayesian analysis in which FAST timing, LAMOST \citep{LAMOST, LAMOST2, LAMOST3} time-domain spectroscopy, and \textit{Gaia} astrometry are simultaneously fitted using a common set of masses and three-dimensional orbital parameters. Such a joint analysis should substantially reduce the geometric and mass degeneracies present in any individual data set and ultimately provide a more complete three-dimensional description of the hierarchical triple PSR~J0435$+$3233 system.

The remainder of this paper is organized as follows.
In Section~\ref{sec:SA_HT}, we introduce the SA description of the PSR~J0435$+$3233 hierarchical triple, following a brief overview of \citetalias{YangHT}. Section~\ref{sec:FAST_HT} incorporates the SA dynamics into the pulsar-timing framework, while Section~\ref{sec:forecast} presents forecasts for distinguishing the Tv1 and Tv2 solutions with continued FAST timing. Sections~\ref{sec:LAMOST_HT} and~\ref{sec:GAIA_HT} discuss the complementary spectroscopic and astrometric prospects, respectively. Section~\ref{sec:discussion} summarizes our main findings and outlines ongoing and planned follow-up efforts.

\section{Singly-Averaged Hierarchical Triple Secular Equations and Their Observational Implications}
\label{sec:dynamics}

We begin by summarizing the innovative efforts of \citetalias{YangHT} that firmly established PSR~J0435$+$3233 as a HT system. We then introduce our SA formulation, tailored to the dynamics of this unique HT system, and explore its implications for the analysis of FAST timing data.

\subsection{\citet{YangHT}'s Description for PSR~J0435$\mathit+$3233 }
\label{sec:Yang_HT}

The pioneering work of \citetalias{YangHT} established the critical physical foundation for analyzing the PSR~J0435$+$3233 triple system by utilizing the local, phenomenological \texttt{ELL1H} binary timing model \citep{ELL1H} to capture rapid, high-order polynomial variations in the pulsar's spin and orbital parameters.

Their mapping is deeply rooted in the analytical framework established by \citet{JR97} for the inversion of high-order pulse frequency derivatives in long-period binary and triple systems. \citet{JR97} demonstrated that when the outer orbital period is much longer than the timing baseline ($P_{\rm orb2} \gg 1$\,yr), standard Keplerian fitting is mathematically nonviable because the data cover only a small fraction of the orbit. However, by measuring up to five successive derivatives of the pulse frequency ($\dot{\nu}$ through $\nu^{(5)}$), the companion mass and the three-dimensional Keplerian parameters of the outer orbit can be completely and uniquely determined up to the standard $\sin i_2$ inclination factor. This is because these derivatives map directly to the line-of-sight components of the pulsar's acceleration ($\mathbf{a} \cdot \hat{\mathbf n}$), jerk ($\dot{\mathbf{a}} \cdot \hat{\mathbf n}$), snap ($\ddot{\mathbf{a}} \cdot \hat{\mathbf n}$), and higher-order derivatives of motion. Under this framework, \citetalias{YangHT} successfully mapped their fitted fifth-derivative spin-down solution, as evident from Table~1 of \citetalias{YangHT} to the orbital characteristics of the G-type subgiant companion.

Crucially, the $\sim99\%$ agreement between the fractional spin and inner-orbital frequency derivatives, $\dot{\nu}/\nu$ and $\dot{f}_{\rm orb1}/f_{\rm orb1}$, respectively, provides strong support for the foundational assumption of the \citet{JR97} model. In other words, \citet{JR97} demands that both quantities are dominated by the same line-of-sight acceleration of the inner binary induced by the distant tertiary, rather than by the intrinsic spin-down of the pulsar. Remarkably, this correspondence extends to higher time derivatives. The measured second derivatives give the following
\begin{equation}
 \frac{\ddot{\nu}/\nu}
      {\ddot{f}_{\rm orb1}/f_{\rm orb1}}
 \simeq 0.97\,,
 \label{eq:nuddot}
\end{equation}
while the corresponding ratio constructed from the third derivatives is
\begin{equation}
 \frac{\nu^{(3)}/\nu}
      {f_{\rm orb1}^{(3)}/f_{\rm orb1}}
 \simeq 0.93\,.
 \label{eq:nudddot}
\end{equation}
Thus, not only the apparent acceleration, but also its higher-order temporal variations due to the changing acceleration (``jerk'') and its derivative (``snap'') along the outer orbit are imprinted coherently on two independent clocks present in the ToAs, namely  the pulsar spin and the inner-binary orbital motion. This hierarchy of matching fractional derivatives provides compelling evidence that the unusually rapid apparent spin evolution of PSR~J0435$+$3233 is predominantly kinematic and originates from its motion in the gravitational field of the tertiary.

In addition to the spin derivatives, \citetalias{YangHT} detected significant time variations in the inner orbit's projected  pulsar semi-major axis ($\dot{x}_1, \ddot{x}_1$) and orbital frequency ($\dot{f}_{\rm orb1}, \ddot{f}_{\rm orb1}, \dddot{f}_{\rm orb1}$). To model these non-Keplerian variations, they incorporated the three-body secular perturbation equations derived by \citet{Rasio1994, JR97}. Specifically, they modeled the secular evolution of the inner orbital inclination angle ($\dot{i}^t_1$) driven by the tertiary star, specified by Eq.~(20) in \citetalias{YangHT}, which leads to temporal variation in the projected pulsar semi-major axis of the inner binary. It turns out that such a perturbation leads to temporal variations in the orbital frequency of the inner binary that can naturally explain the deduced derivatives of ${f}_{\rm orb1}$ as listed in Table~1 of \citetalias{YangHT}. Incidentally, these three-body perturbations also produce temporal variations to inner orbit eccentricity and periastron precession, although \citetalias{YangHT} neglected these evolutions. By integrating these effects into the pulsar timing software \texttt{Tempo2} \citep{Tempo2, Tempo2II}, \citetalias{YangHT} successfully modeled the measured ToAs from FAST, which allowed them to map out the system's true three-dimensional geometries and identify the currently degenerate solutions, namely Tv1 and Tv2. The two viable solutions have very similar orbital properties but differ appreciably in their masses and three-dimensional geometry. Both favor an $\simeq8$-day, nearly circular ($e_1\simeq1.6\times10^{-4}$) inner orbit and an eccentric ($e_2\simeq0.60$), $\simeq74$-yr outer orbit. Tv1 favors masses $(M_{\rm PSR},M_{\rm WD},M_\star)\simeq(1.15,0.27,0.96)\,M_\odot$ and a mutual inclination $i_X\simeq84^\circ$, whereas Tv2 favors $(M_{\rm PSR},M_{\rm WD},M_\star)\simeq(1.29,0.30,1.12)\,M_\odot$ and $i_X\simeq55^\circ$. In other words, the present degeneracy is not primarily between substantially different orbital periods or eccentricities, but between different mass scales and three-dimensional orientations of the HT.
  
Building directly upon this breakthrough, we translate \citetalias{YangHT}'s epoch-dependent polynomial description into  a singly averaged  formulation that averages over the rapid inner orbital motion while retaining the instantaneous outer orbital phase, thereby capturing the time-dependent quadrupolar perturbations exerted by the tertiary on the inner binary.

\subsection{ Singly Averaged Vectorial Equations for HT PSR~J0435$\mathit +$3233  }
\label{sec:SA_HT}

The substantial eccentricity of the outer orbit causes the tidal perturbation exerted by the tertiary on PSR~J0435$+$3233 binary to vary strongly over an outer orbital cycle (\citetalias{YangHT}). A local Taylor-series description in terms of successive time derivatives of the inne orbital elements can provide an excellent representation over the presently available timing baseline. However, its predictive power is expected to deteriorate as the system approaches the outer periastron passage in 2036, where the rapidly decreasing separation $R$ enhances the quadrupolar perturbation approximately as $R^{-3}$, as evident from Eq.~\eqref{eq:QH}. The resulting strongly time-dependent evolution of the inner orbit becomes increasingly difficult to represent with a finite set of constant higher-order derivatives, requiring progressively more terms and introducing additional covariances among the fitted parameters. This provides a strong motivation for replacing such a local phenomenological expansion with a SA formulation of HT dynamics, where we only average over the rapid motion of the inner binary, while retaining the instantaneous position of the tertiary.

Influenced by \citet{Luo2016, Toonen2016,Hamers2020}, the singly averaged quadrupolar perturbation term in the Hamiltonian per unit reduced mass reads 
\begin{equation}
    \Phi_{\rm SA} = \frac{G M_{\star} a_{\rm in}^2}{4 R^3} \left[ -1 + 6e_1^2 + 3(\mathbf{j}_1 \cdot \hat{\mathbf R})^2 - 15(\mathbf{e}_1 \cdot \hat{\mathbf R})^2 \right]\,,
 \label{eq:QH}
\end{equation}
where $R$ is the instantaneous physical distance between the tertiary companion star and the center of mass (barycenter) of the inner binary while $a_{\rm in}$ stands for the relative semi-major axis of the inner binary. The secular state of the inner binary is specified by its eccentricity vector $\mathbf{e}_1$ and dimensionless angular-momentum vector $\mathbf{j}_1~=~\sqrt{1-e_1^2}\,\hat{\mathbf j}_1$(the dimensionless Laplace-Runge-Lenz vector). Additionally, the instantaneous separation $R (t)$ is specified in terms of its Newtonian Keplerian parametric solution as $R(t) = a_{\rm out}(1 - e_2^2)/(1 + e_2\cos f_2)$, where $a_{\rm out}, e_2$ and $f_2$ stand for the semi-major axis, eccentricity and the true anomaly of the outer orbit while $\hat{\mathbf R}(t)$ is the unit vector pointing along the outer relative separation vector. Further, $M_{\star}$ stands for the mass of the tertiary while $a_{\rm in}$ represents the semi-major axis of the relative inner orbit consisting of the pulsar and the white dwarf. The SA model for our HT  is uniquely optimal because it retains this outer phase dependence without the computational burden of full, non-averaged 3-body integrations \citep{WisdomNbody} over thousands of inner orbits. We note in passing that the dimensionless angular-momentum vector $\mathbf{j}_1$ and the eccentricity vector $ \mathbf e_1 = e_1\hat{\mathbf e}_1 $ are defined such that 
\begin{equation}
\mathbf e_1\cdot\mathbf j_1=0\,,
\qquad
e_1^2+j_1^2=1\,.
\label{eq:ejvec}
\end{equation}

We describe the secular evolution of the inner binary containing PSR~J0435$+$3233 using the Milankovitch vectorial formulation of celestial mechanics \citep{Milankovitch1939, Tremaine2009}, which provides a particularly convenient description of hierarchical three-body dynamics \citep{ValtonenKarttunen2006}. The relevant equations are 
\begin{subequations}
\label{eqs:Mil}
\begin{align}
    \frac{\mathrm{d}\mathbf{j}_1}{\mathrm{d}t}
    &=
    -\frac{1}{\Lambda_1}
    \left[
    \mathbf{j}_1 \times
    \frac{\partial\Phi_{\rm SA}}{\partial\mathbf{j}_1}
    +
    \mathbf{e}_1 \times
    \frac{\partial\Phi_{\rm SA}}{\partial\mathbf{e}_1}
    \right],
    \label{eq:MilJ}
    \\
    \frac{\mathrm{d}\mathbf{e}_1}{\mathrm{d}t}
    &=
    -\frac{1}{\Lambda_1}
    \left[
    \mathbf{e}_1 \times
    \frac{\partial\Phi_{\rm SA}}{\partial\mathbf{j}_1}
    +
    \mathbf{j}_1 \times
    \frac{\partial\Phi_{\rm SA}}{\partial\mathbf{e}_1}
    \right].
    \label{eq:MilE}
\end{align}
\end{subequations}
where $ \Lambda_1 = \sqrt{GM_{\rm in}a_{\rm in}} $, with $ M_{\rm in} \equiv M_{\rm PSR} + M_{\rm WD}$ being the total mass of the inner binary. Defining the characteristic perturbation frequency as $\nu_{\rm SA}~\equiv~G M_{\star} a_{\rm in}^2 / (\Lambda_1 R^3)$, we obtain the following secular coupled  ordinary differential equations for $\mathbf{j}_1$ and $\mathbf{e}_1$:
\begin{subequations}
\label{eqs:djde}
\begin{align}
    \frac{\mathrm{d}\mathbf{j}_1}{\mathrm{d}t} &= \nu_{\rm SA} \left[ -\frac{3}{2}(\mathbf{j}_1 \cdot \hat{\mathbf R})(\mathbf{j}_1 \times \hat{\mathbf R}) + \frac{15}{2}(\mathbf{e}_1 \cdot \hat{\mathbf R})(\mathbf{e}_1 \times \hat{\mathbf R}) \right]\,,
    \label{eq:djdt}
    \\
    \frac{\mathrm{d}\mathbf{e}_1}{\mathrm{d}t} &= \nu_{\rm SA} \left[ -3(\mathbf{j}_1 \times \mathbf{e}_1) - \frac{3}{2}(\mathbf{j}_1 \cdot \hat{\mathbf R})(\mathbf{e}_1 \times \hat{\mathbf R}) \right. \nonumber \\
    &\quad \quad \quad \left. + \frac{15}{2}(\mathbf{e}_1 \cdot \hat{\mathbf R})(\mathbf{j}_1 \times \hat{\mathbf R}) \right] + \dot \omega_{\rm GR} (\hat{\mathbf{j}}_1 \times \mathbf{e}_1)\,.
    \label{eq:dedt}
\end{align}
\end{subequations}

The second term in Eq.~\eqref{eq:dedt} represents the first order post-Newtonian (1PN) general relativistic (GR) periastron precession, and it produces only an apsidal rotation of the eccentricity vector $\mathbf e_1$, such that 
\begin{equation}
\dot{\mathbf e}_1
=
\dot{\mathbf e}_{1,\rm SA}
+
\dot{\mathbf e}_{1,\rm GR}\,,
\label{eq:e1dot}
\end{equation}
where the subscripts SA and GR represent the contributions from the SA dynamics and the general relativistic periastron precession respectively. The 1PN-accurate expression for the rate of periastron precession reads \citep{Damour1992} 
\begin{equation}
\dot \omega_{\rm GR}
=
\frac{3n_{\rm in}GM_{\rm in}}
{a_{\rm in}c^2(1-e_1^2)}\,,
\label{eq:omegadotgr}
\end{equation}
where $ n_{\rm in} = \sqrt{{GM_{\rm in}}/{a_{\rm in}^3}} $ gives the mean motion of the inner binary. This contribution is dynamically critical as the inner relativistic precession timescale ($1/\dot \omega_{\rm GR} \approx 7 \times 10^3\,\rm yr$) is much shorter than the Zeipel-Lidov-Kozai (ZLK) timescale \citep{Blaes2002, Naoz2016} ($t_{\rm ZLK} \approx 5 \times 10^4\,\rm yr$), the fast GR precession completely quenches the ZLK eccentricity oscillations \citep{Liu2015}. Consequently, the inner orbit remains strictly near-circular ($e_1 \approx 10^{-4}$) over long-term integration, preventing catastrophic eccentricity growth (\citetalias{YangHT}).

A few comments are in order. Note that  $\mathbf{R}(t) = R(t)\, \hat{\mathbf R}(t) $ provides the instantaneous position of the tertiary relative to the inner-binary barycenter. Interestingly,  this variable needs not be included as an additional dynamical variable in the six-dimensional SA system.  Instead, at each integration step, $\mathbf{R}(t)$ can be reconstructed from the outer Keplerian orbit specified by the typical six Newtonian parameters, required to uniquely identify, orient, and size an unperturbed orbit in three-dimensional space, given by \citep{Murray1999}:
\begin{equation}
    \left(
    P_2,e_2,T_{0,2},\omega_2,i_2,\Omega_2
    \right)\,.
    \label{eq:outerorbitparams}
\end{equation}

The instantaneous position of the tertiary relative to the inner-binary barycenter is obtained by solving its Kepler equation which requires us to define an outer mean anomaly $l_2$ as
\begin{equation}
    l_2(t)
    =
    \frac{2\pi}{P_2}
    \left(t-T_{0,2}\right)\,.
    \label{eq:meananomaly}
\end{equation}
Thereafter, the eccentric anomaly $E_2$ follows from the Kepler equation,
\begin{equation}
    l_2=E_2-e_2\sin E_2\,,
    \label{eq:keplereqn}
\end{equation}
while the true anomaly $f_2$ is subsequently obtained through
\begin{equation}
    \tan\frac{f_2}{2}
    =
    \sqrt{\frac{1+e_2}{1-e_2}}
    \tan\frac{E_2}{2}\,.
    \label{eq:trueanomaly}
\end{equation}

With these inputs,  the instantaneous outer orbit separation $R(t)$ is
\begin{equation}
    R(t)
    =
    a_{\rm out}\left(1-e_2\cos E_2\right)\,
    \label{eq:Rt}
\end{equation}
while the corresponding unit vector $\hat{\mathbf R}(t)$ in the observer frame is given by 
\begin{equation}
\hat{\mathbf{R}}(t)
=
\begin{pmatrix}
\cos\Omega_2 \cos(\omega_2+f_2)
-\sin\Omega_2 \sin(\omega_2+f_2)\cos i_2
\\[2mm]
\sin\Omega_2 \cos(\omega_2+f_2)
+\cos\Omega_2 \sin(\omega_2+f_2)\cos i_2
\\[2mm]
\sin(\omega_2+f_2)\sin i_2
\end{pmatrix}\,.
\label{eq:Rhat}
\end{equation}

Clearly, we need to specify the initial conditions for the six SA coupled ordinary differential equations, given by Eqs.~\eqref{eqs:djde}, and the required initial state vector at some reference epoch $t_0$ reads 
\begin{equation}
\mathbf{y}(t_0)
=
\left(
j_x,j_y,j_z,e_x,e_y,e_z
\right)_{t_0}\,.
\label{eq:y_t0}
\end{equation}

The advantage of the SA formalism is that the timing results, available in Table~4 of \citetalias{YangHT}, could be employed to specify the temporal evolution of our system of interest: PSR~J0435$+$3233. For example, the initial eccentricity vector can be written directly in terms of the Laplace--Lagrange parameters of the inner binary, namely $ \epsilon_1=e_1\sin\omega_1$, $\epsilon_2=e_1\cos\omega_1 $, as
\begin{equation}
    \mathbf{e}_{1,0}
    =
    \begin{pmatrix}
        \epsilon_2\\
        \epsilon_1\cos i_1\\
        \epsilon_1\sin i_1
    \end{pmatrix} \,,
    \label{eq:initial_evec}
\end{equation}
where we have exploited the freedom to rotate the coordinate system around the line of sight and choose $\Omega_1 (t_0)=0$. The use of the Laplace--Lagrange parameters implies that 
\begin{equation}
    e_1
    =
    \sqrt{\epsilon_1^2+\epsilon_2^2}\,,
    \label{eq:e1mag}
\end{equation}
and
\begin{equation}
    \omega_1
    =
    \operatorname{atan2}(\epsilon_1,\epsilon_2)\,.
    \label{eq:omega1}
\end{equation}
Similarly, the initial dimensionless angular-momentum vector becomes
\begin{equation}
    \mathbf{j}_{1,0}
    =
    \sqrt{1-e_1^2}
    \begin{pmatrix}
        0\\
        -\sin i_1\\
        \cos i_1
    \end{pmatrix}\,.
    \label{eq:initial_jvec}
\end{equation}

Interestingly, the SA integration directly provides the temporal evolution of the orientation of the inner orbital plane, as we are allowed to let the $z$-axis to coincide with the line of sight. This ensures that the inner binary's inclination becomes
\begin{equation}
    \cos i_1(t)
    =
    \frac{j_{1,z}(t)}
         {|\mathbf{j}_1(t)|}\,.
    \label{eq:i1t}
\end{equation}
This implies that the SA integration allows us to track the geometrical projected semi-major axis $a_1$of the pulsar around the inner binary barycenter: 
\begin{equation}
    x_{1,\rm SA}(t)
    =
    \frac{a_{1}}{c}\sin i_1(t)\,.
    \label{eq:xSAt1}
\end{equation}
Using the measured projected semi-major axis $x_{1,0}$ at the reference epoch $t_0$, this can be conveniently written as
\begin{equation}
    x_{1,\rm SA}(t)
    =
    x_{1,0}
    \frac{\sin i_1(t)}
         {\sin i_1(t_0)}\,.
    \label{eq:xSAt2}
\end{equation}
Equivalently, it can be calculated directly from the integrated angular-momentum vector:
\begin{equation}
    x_{1, \rm SA}(t)
    =
    x_{1,0}
    \frac{
    \sqrt{1-j_{1,z}^2(t)/j_1^2(t)}
    }{
    \sqrt{1-j_{1,z}^2(t_0)/j_1^2(t_0)}
    }\,.
    \label{eq:xSAt}
\end{equation}
The corresponding nodal contribution to the first derivative is
\begin{equation}
    \dot{x}_{1, \rm SA}
    =
    x_{1, \rm SA}\cot i_1\,\dot{i}_1\,.
    \label{eq:xSAdot}
\end{equation}

In other words, $\epsilon_1(t), \epsilon_2(t), x_{1, \rm SA}(t)$ and $\dot{x}_{1,\rm SA}(t)$ follow directly from the numerical solution of the SA equations, given by Eqs.~\eqref{eqs:djde}.

\begin{figure*}
    \centering
    \includegraphics[width=0.90\textwidth]{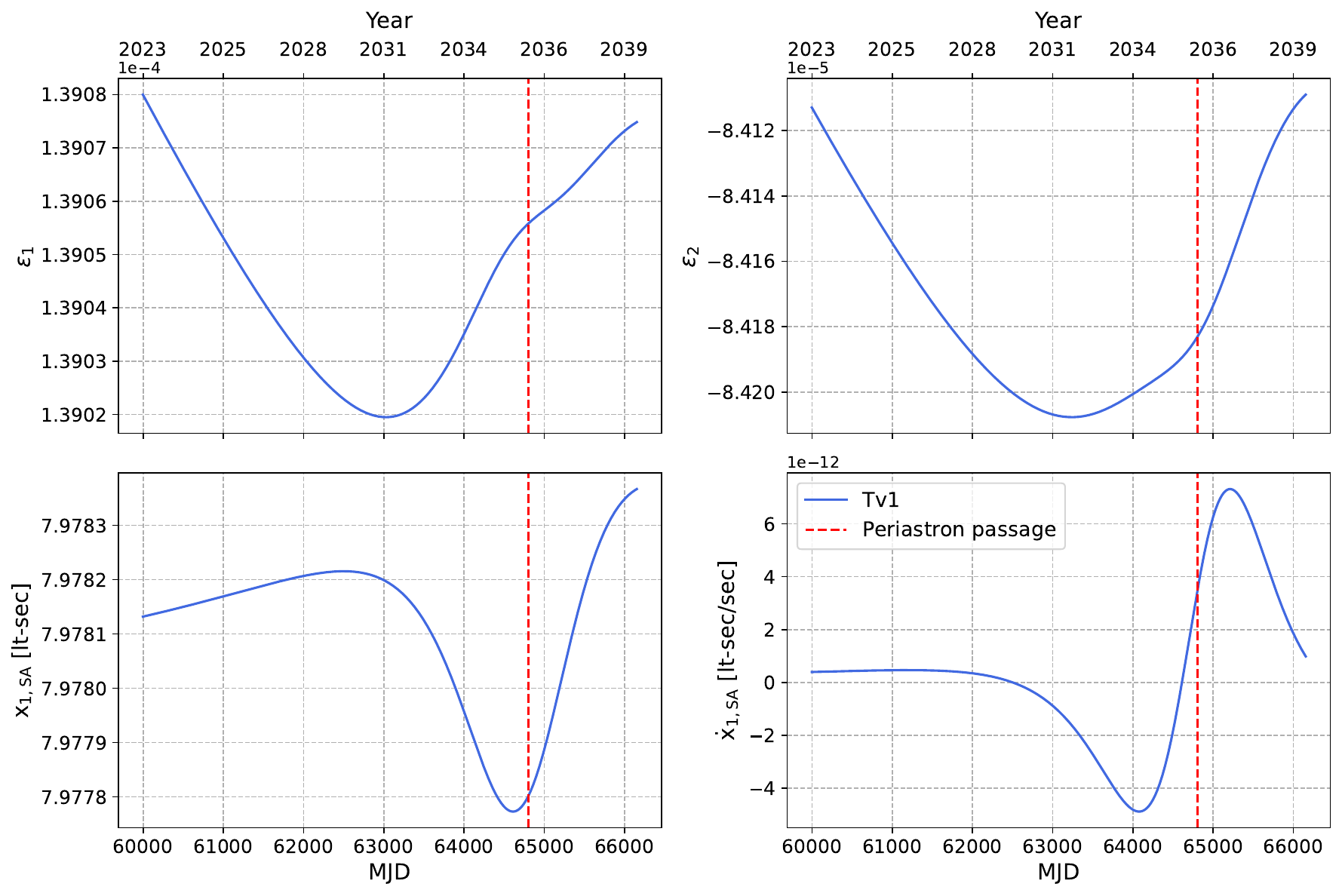}
    \caption{Time evolution of the inner-orbit dynamical variables under the SA dynamics, initialized using the Tv1 timing solution listed in Table~4 of \citetalias{YangHT}. The top panels display the Laplace--Lagrange parameters, $\epsilon_1=e_1\sin\omega_1$ and $\epsilon_2=e_1\cos\omega_1$, while the bottom panels show the projected semi-major axis, $x_{1,\rm SA}$, and its time derivative, $\dot{x}_{1,\rm SA}$. The vertical dashed line marks the periastron passage of the outer orbit, with the corresponding calendar years indicated along the upper axes. Notably, the predicted changes in all four quantities over the coming years substantially exceed the current measurement uncertainties of their corresponding timing observables, suggesting that the SA-driven evolution should be directly detectable with continued FAST observations.}
    \label{fig:sa_observables}
\end{figure*}

We now initialize our SA framework using the parameters listed in Table~4 of \citetalias{YangHT} and evolve the inner-orbit observables $\epsilon_1(t)$, $\epsilon_2(t)$, $x_{1, \rm SA}(t)$, and $\dot{x}_{1, \rm SA}(t)$. This allows us to track the response of these dynamical variables to the strongly varying tidal influence of the tertiary along its eccentric outer orbit, particularly during the approach to periastron and the results are displayed in Fig.~\ref{fig:sa_observables}.

As expected, we find pronounced variations in $x_{1, \rm SA}(t)$ and $\dot{x}_{1, \rm SA}(t)$ as the tertiary approaches and passes through periastron, reflecting the rapidly strengthening three-body perturbation. More importantly, the predicted evolution of all four dynamical variables over the coming years is substantially larger than the current measurement uncertainties of their corresponding timing observables. This suggests that the secular evolution of these observables should  become directly measurable with continued FAST timing, well before the outer periastron passage. This provides a strong observational motivation for incorporating our SA formulation directly into the timing analysis of PSR~J0435$+$3233, allowing the measured evolution to be described by the underlying three-body dynamics rather than through a succession of phenomenological higher-order derivatives.

It is useful to contrast the present SA formulation with the doubly averaged (DA) treatment commonly employed to investigate the long-term secular evolution of HTs \citep{Gopakumar2009,Shukla2025}. For example, one of us recently employed a quadrupolar-order DA description to constrain the proposed tertiary companion to the eccentric MSP binary PSR~J1618$-$3921 \citep{Shukla2025}. In the DA approximation, the perturbing Hamiltonian is averaged over the mean anomalies of both the inner and outer orbits. Consequently, the explicit dependence on the instantaneous outer orbital phase is removed, yielding equations that describe the long-term secular evolution of quantities such as the inner eccentricity and argument of periastron. Such an approach is well suited to probing the cumulative influence of a very wide tertiary when its orbital phase is poorly constrained and the observational quantities of interest are predominantly secular.

The same approximation is not appropriate for our present purpose in PSR~J0435$+$3233. Its eccentric outer orbit ($e_2\simeq0.60$) causes the strength of the tertiary perturbation to vary strongly around the $\simeq74$-yr orbit, becoming particularly pronounced during the approach to the periastron passage around mid-2036. Averaging over the outer anomaly would replace this strongly phase-dependent interaction by its orbit-averaged value. Therefore, this procedure would erase precisely the short-timescale outer-orbit modulation that we seek to predict with continued FAST timing. In particular, the observed evolution of $\mathbf{e}_1$ and $\mathbf{j}_1$, and hence of $\epsilon_1$, $\epsilon_2$, $i_1$, and $x_{1, \rm SA}$, depends on the instantaneous separation and orientation of the tertiary through $R(t)$ and $\hat{\mathbf R}(t)$. The SA formulation provides the appropriate intermediate description: it averages over the rapid $\simeq8$-day inner orbit while retaining the instantaneous outer orbital phase, thereby preserving the strongly time-dependent quadrupolar forcing while avoiding the computational expense of a fully unaveraged three-body integration.

\subsection{Probing Pulsar Timing Implications}
\label{sec:FAST_HT}

In the pioneering effort of \citetalias{YangHT}, the presence of tertiary in the FAST ToAs of PSR~J0435$+$3233 was extracted using the \texttt{ELL1H} timing model \citep{ELL1H}. This model is a direct high-precision extension of the \texttt{ELL1} timing model \citep{ELL1}. While \texttt{ELL1} elegantly handles low-eccentricity orbits by replacing $e$ and $\omega$ with the Laplace--Lagrange parameters $( \epsilon_1, \epsilon_2)$, it does not solve the severe parameter covariance issues that arise when attempting to measure the Shapiro delay in circular systems. The \texttt{ELL1H} model was designed specifically to break this remaining degeneracy using certain orthometric parameterization \citep{ELL1H}.

In the standard \texttt{ELL1H} binary model, the primary geometric timing delay is the  R{\o}mer delay $\Delta_{\rm R}(t)$ which is written for low-eccentricity orbits to first order in eccentricity $e_1$ as \citep{ELL1H}:
\begin{align}
\Delta_{\rm R1}(t) &= x_1(t) \sin \Phi_1(t) - \frac{x_1(t)}{2} \Bigl [ \epsilon_1(t) \cos 2\Phi_1(t) 
\nonumber\\
&\hspace{2.6cm}
- \epsilon_2(t) \sin 2\Phi_1(t) \Bigr ]\,,
\label{eq:roemerell1h}
\end{align}
where $x_1(t)$ is the projected semi-major axis of the pulsar's inner orbit, while the mean orbital phase is propagated linearly from the reference epoch of ascending node passage $T_{\rm{asc1}}$ as:
\begin{equation}
    \Phi_1(t) = 2\pi f_{\rm{orb1}} (t - T_{\mathrm{asc1}})\,.
    \label{eq:phasepulsar}
\end{equation}
In \citetalias{YangHT}, the Laplace--Lagrange parameters $\epsilon_1$ and $\epsilon_2$ are fitted as static constants, while both $x_1(t)$ and $ f_{\mathrm{orb1}}(t)$ are modeled as a local phenomenological polynomial expansion. This allows one to express, for example, $x_1(t)$ as 
\begin{equation}
    x_1(t) = x_{1,0} + \dot{x}_1 (t - t_0) + \frac{1}{2}\ddot{x}_1 (t - t_0)^2 + \dots\,.
    \label{eq:xt_pheno}
\end{equation}
In the presence of the active tidal torque from the tertiary star, both $ x_1$ and $f_{\mathrm{orb1}}$ are time-varying, as evident in Table~1 of \citetalias{YangHT}. In other words, the time-dependent evolution of the inner orbit is captured with the help of  a finite set of constant higher-order derivatives.
 
To upgrade the phenomenological \texttt{ELL1H} model into a self-consistent physical model, we propose to replace the static or polynomial parameters with the instantaneous trajectories computed from the SA secular equations. Influenced by \citetalias{YangHT}, we incorporate the effects of classical Doppler and transverse-Doppler kinematic corrections induced by the inner binary barycenter's motion in the outer orbit while evaluating the R\"omer delay. This leads to an effective projected semi-major axis $x_{1,\rm{eff}}(t)$ as: 

\begin{equation}
    x_{1,\mathrm{eff}}(t) = x_{1,\mathrm{SA}}(t) \left[ 1 - \frac{v_{2,\parallel}(t)}{c} \right] + \frac{a_1}{c} \frac{v_{2,y}(t)}{c} + C \,,
    \label{eq:xeff}
\end{equation}
where $v_{2,\parallel}(t) = \mathbf{v}_2(t) \cdot \hat{\mathbf z}$ is the line-of-sight velocity of the inner binary's barycenter in the outer orbit which provides the classical Doppler contribution. The second contribution requires us to compute $v_{2,y}(t) = \mathbf{v}_2(t) \cdot \hat{\mathbf{q}}_1(t)$ and it provides the component of the barycentric velocity lying in the instantaneous inner orbital plane and perpendicular to the inner ascending node. The relevant unit vector $\hat{\mathbf{q}}_1(t)$ is computed from the integrated states as $\hat{\mathbf{q}}_1(t) = \hat{\mathbf{j}}_1(t) \times \hat{\mathbf{N}}_1(t)$, where $\hat{\mathbf{N}}_1(t) = (\hat{\mathbf z} \times \hat{\mathbf{j}}_1(t))/|\hat{\mathbf z} \times \hat{\mathbf{j}}_1(t)|$. Further, the constant $C$ ensures that we could align  the integrated profile to the measured timing coordinate $x_{1,0}$ at the reference epoch $t_0$. Armed with these inputs, and influenced by \citetalias{YangHT} we write the SA-updated  effective R{\o}mer delay as

\begin{align}
    \Delta_{\mathrm{R1, eff}}(t) &= \frac{x_{1,\mathrm{eff}}(t)}{c} 
    \Bigl [ 
    \sin\Phi_{1,\mathrm{eff}}(t) + \frac{1}{2} \epsilon_2(t) \sin 2\Phi_{1,\mathrm{eff}}(t)
    \nonumber\\
&
    \quad \quad \quad \quad \quad - \frac{1}{2} \epsilon_1(t) \cos 2\Phi_{1,\mathrm{eff}}(t) 
    \Bigr ]\,,
    \label{eq:roemer_ell11_full}
\end{align}
where $\Phi_{1, \mathrm{eff}}(t) = \Phi_{1}(t) + (v_{2,x}(t)/c \sin i_1(t))$ with  $  v_{2,x}(t) = \mathbf{v}_2(t) \cdot \hat{\mathbf N}_1(t)$. 

An additional subtlety needs to be addressed while converging on an appropriate R{\o}mer delay expression for the inner binary. In the phenomenological \texttt{ELL1H} timing transformation, as given by Eq.~(25) in \citetalias{YangHT}, the inner R\"omer delay $\Delta_{\rm R1}(t)$ was evaluated omitting certain non-periodic term $-\frac{3}{2}\frac{x_1, \rm eff}{c}\epsilon_1(t)$, as noted in \citet{Susobhanan2018}. Typically, this term is absorbed into a constant reference phase and a linear spin-frequency shift  under  purely linear evolution \citep{Susobhanan2018}. However, it has a significant baseline magnitude of $\approx 1.67\,\mathrm{ms}$ for PSR~J0435$+$3233 ($x_1 \approx 7.978\,\mathrm{lt\text{-}s}, \epsilon_1 \approx 1.39 \times 10^{-4}$). Further, as the tertiary companion approaches outer periastron, three-body secular perturbations drive non-linear variations in $\epsilon_1(t)$, causing this term to generate non-linear timing drifts that cannot be absorbed by a constant frequency shift. Therefore, we employ the following expression for R\"omer delay, influenced by \citet{Susobhanan2018}:
\begin{align}
    \Delta_{\mathrm{R1, eff}}(t) &= \frac{x_{1,\mathrm{eff}}(t)}{c} 
    \Bigl [ 
    \sin\Phi_{1,\mathrm{eff}}(t) 
    + \frac{1}{2} \epsilon_2(t) \sin 2\Phi_{1,\mathrm{eff}}(t)
       \nonumber\\
& 
    \quad - \frac{1}{2} \epsilon_1(t) \cos 2\Phi_{1,\mathrm{eff}}(t) 
    \Bigr ]
     - \frac{3}{2} \frac{x_{1,\mathrm{eff}}(t)}{c} \epsilon_1(t)\,.
    \label{eq:roemer_ell1k_full}
\end{align}

\begin{figure*}
    \centering
    \includegraphics[width=0.90\textwidth]{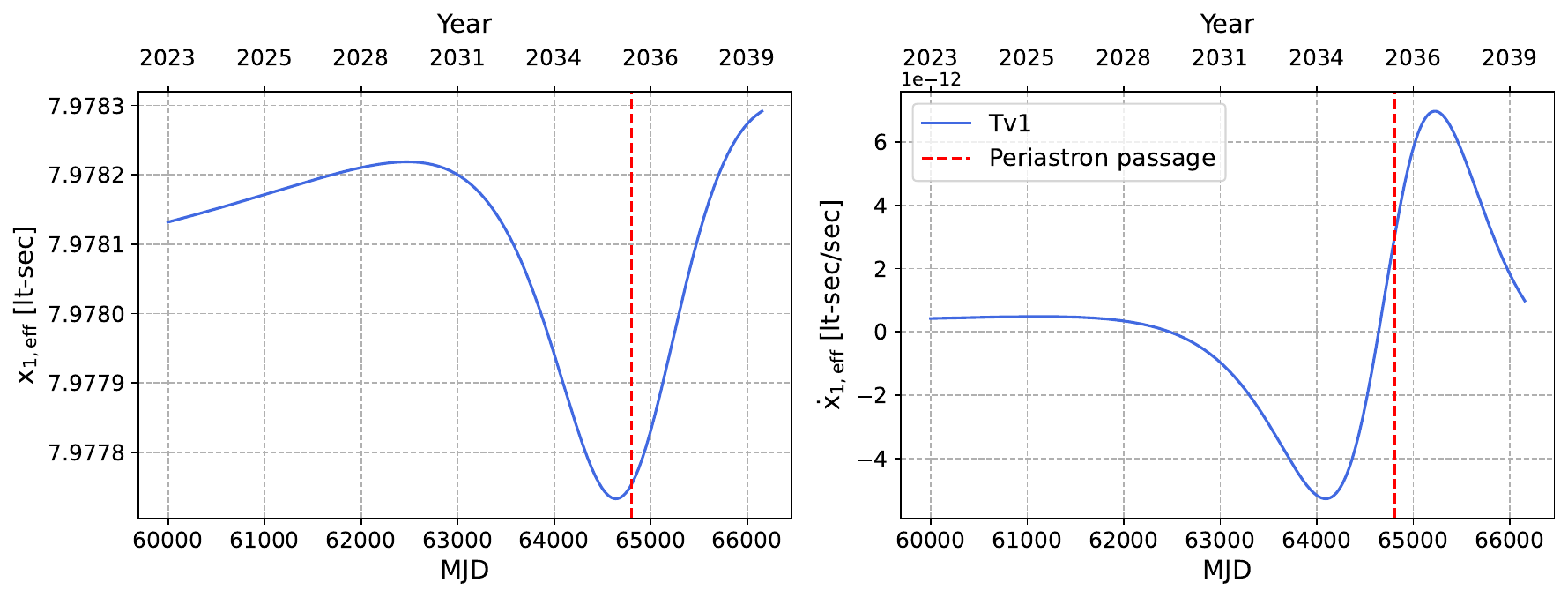}
    \caption{Time evolution of the effective projected semi-major axis of the inner binary $x_{1, \rm eff}$ (left panel) and its time derivative $\dot{x}_{1, \rm eff}$ (right panel), initialized using the Tv1 timing solution listed in Table~4 of \citetalias{YangHT}. The vertical dashed line marks the periastron passage of the outer orbit. The time evolution of these observables remain qualitatively similar to the SA evolution (Fig.~\ref{fig:sa_observables}), as the additional kinematic contributions are suppressed by a factor of order $v_2/c$, which is typically $\sim10^{-6}$--$10^{-5}$ for this system. Nevertheless, these terms are important because their magnitude is potentially measurable with the timing precision of FAST and, more fundamentally, because they provide a complete physical description of the system by accounting for the motion of the inner binary's center of mass.}
    \label{fig:xeff_observables}
\end{figure*}

Therefore, the only difference between Eq.~\eqref{eq:roemer_ell11_full} and Eq.~\eqref{eq:roemer_ell1k_full} is the explicit retention of the $-\frac{3}{2}\frac{x_{1,\mathrm{eff}}}{c}\epsilon_1(t)$ contribution, as appropriate for  PSR~J0435$+$3233. We note in passing that we are allowed to employ $x_{1, \mathrm{eff}}$  and $ \Phi_{1, \mathrm{eff}} $ in the last three terms due to the presence of subdominant Laplace--Lagrange parameters in these contributions. In Fig.~\ref{fig:xeff_observables}, we plot the temporal evolution of $ x_{1, \mathrm{eff}}(t) $ and its time-derivative and we observe that their time evolution remain qualitatively similar to the SA evolution, as displayed in  Fig.~\ref{fig:sa_observables}. This is not surprising as the additional corrections are
of order $v_2/c$.

Another relativistic ingredient of the \texttt{ELL1H} timing model, employed by \citetalias{YangHT} in their analysis of PSR~J0435$+$3233, is the Shapiro propagation delay. In high-precision pulsar timing, this general-relativistic delay arises as the pulsar signal propagates through the gravitational field of its companion and provides a powerful probe of the companion mass and orbital inclination \citep{DD86}. Because the inner orbit of PSR~J0435$+$3233 is extremely circular ($e_1 \approx 0.00016$), standard relativistic parameters (such as the range $r$ and shape $s = \sin i_1$) exhibit severe mutual covariances with the Keplerian R{\o}mer delay \citep{ELL1, ELL1H}. To eliminate these degeneracies, \citetalias{YangHT} employed the \texttt{ELL1H} binary model \citep{ELL1H}, parameterizing the inner Shapiro delay ($\Delta_{\rm S1}$) via the orthometric parameters as noted earlier. These parameters, introduced in \citet{ELL1H}, are the third orthometric amplitude ($h_3$) and the orthometric ratio ($\varsigma$). 

In general relativity, these orthometric parameters are related to the physical properties of the inner helium WD companion by \citep{ELL1H}:
\begin{subequations}
\begin{align}
    h_3 &= \frac{G M_{\rm WD}}{c^3} \varsigma^3\,, \label{eq:h3_def} \\
    \varsigma &= \frac{\sin i_1}{1 + |\cos i_1|}\,. \label{eq:varsigma_def}
\end{align}
\end{subequations}
An important advantage of our SA description of the HT is that it naturally provides the time-dependent inner-orbit inclination $i_1(t)$, as evident from Eq.~\eqref{eq:i1t}, thereby allowing the corresponding temporal evolution of the Shapiro delay to be incorporated self-consistently into the timing model.

We note in passing that, when we eventually analyze the FAST ToAs using our SA formalism incorporated into the \texttt{ELL1H} model, the fitted timing parameters will be determined by those required for specifying SA integrations and those appear in various delays. The relevant 16 dimensional parameter set reads 

\begin{widetext}
\begin{equation}
\boldsymbol{\theta}_{\mathrm{SA}} = \Big\{ 
\underbrace{T_{\mathrm{asc1}}, P_{\mathrm{orb1}}, x_1, \epsilon_{1}, \epsilon_{2}, i_1}_{\text{Inner orbit at } t_0}, 
\quad 
\underbrace{P_{\mathrm{orb2}}, x_2, e_2, T_{02}, \omega_2, i_2, \Delta\Omega}_{\text{Outer orbit at } t_0}, 
\quad 
\underbrace{M_{\mathrm{PSR}}, M_{\mathrm{WD}}, M_{\star}}_{\text{Component masses}} 
\Big\}\,.
\label{eq:mcmc_orbital_vector}
\end{equation}
\end{widetext}

It is advisable to retain $x_1$ and $x_2$ as free parameters, since the FAST ToAs should allow them to be constrained to sub-microsecond precision, as demonstrated by \citetalias{YangHT}. The physical semi-major axes required for integrating Eq.~\eqref{eqs:djde} can then be readily computed from these fitted quantities using the following Newtonian relations:
\begin{subequations}
\label{eq:semi_major_axes}
\begin{align}
    a_{\rm in}
    &= \frac{c\,x_1}{\sin i_1}
    \left( \frac{M_{\mathrm{PSR}} + M_{\mathrm{WD}}}{M_{\mathrm{WD}}} \right)\,,
    \label{eq:a_in}
    \\
    a_{\mathrm{out}}
    &= \frac{c\,x_2}{\sin i_2}
    \left( \frac{M_{\mathrm{in}} + M_{\star}}{M_{\star}} \right)\,.
    \label{eq:a_out}
\end{align}
\end{subequations}

We now explore the implications of our approach for  the FAST timing campaign on PSR~J0435$+$3233.

\subsection{Secular Orbital Forecasts for FAST}
\label{sec:forecast}

To forecast when the present geometric degeneracy between solutions Tv1 and Tv2 may become observationally distinguishable, we forward-integrate the coupled SA evolution equations from the reference epoch, $t_0=\mathrm{MJD}\,59997.5$ (2023 February 22), through 2040 January 1, thereby encompassing the crucial 2036 outer periastron passage. Ideally, a statistically robust forecast of the future distinguishability of Tv1 and Tv2 solutions would be obtained through Monte Carlo propagation of samples drawn directly from their respective joint posterior distributions. Such an analysis would preserve the correlations among the masses, orbital elements, and three-dimensional orientation parameters,  evident in the posterior distributions, displayed in Fig.~7 of \citetalias{YangHT}, and would yield time-dependent posterior-predictive uncertainty bands for $x_{1, \rm eff}(t)$ and other timing observables. However, the posterior samples underlying the Tv1 and Tv2 solutions are currently unavailable. Consequently, a fully correlated posterior-predictive forecast cannot be constructed from the published results alone.

We therefore quantify the expected separation of the two solutions using the approximate time-dependent statistic
\begin{equation}
{\cal S}_x(t)=
\frac{
\left|{x}^{\rm Tv1}_{1,\rm eff}(t)-{x}^{\rm Tv2}_{1,\rm eff}(t)\right|
}{
\left[
\sigma^2_{x,\rm Tv1}(t)+
\sigma^2_{x,\rm Tv2}(t)+
\sigma^2_{x,\rm obs}(t)
\right]^{1/2}
},
\label{eq:Sx}
\end{equation}
where $x$ stands for $x_{1,\rm eff}(t)$ while ${x}^{\rm Tv1(2)}_{1, \rm eff}(t)$ denote the predicted values for the two solutions. In addition, $\sigma_{x,\rm Tv1(2)}(t)$ characterize their propagated uncertainties inferred from the published parameter constraints, and $\sigma_{x,\rm obs}(t)$ represents the expected observational uncertainty, and we extracted these uncertainties from Tables~4 and 1 of \citetalias{YangHT}.
 
In our opinion, Eq.~(\ref{eq:Sx}) provides a crucial \emph{model-penalized} distinguishability metric as it adds the expanding model variances ($\sigma^2_{x,\mathrm{Tv1}}(t) + \sigma^2_{x,\mathrm{Tv2}}(t)$) directly in quadrature with the instrumental noise floor ($\sigma^2_{x,\mathrm{obs}}$). This  ensures that a threshold of $S_x(t) \ge 5$ represents an absolute guarantee of statistical distinction by a significance of $5\sigma$. The two trajectories are declared observationally distinguished only when their physical separation exceeds both telescope timing noise \emph{and} the combined parameter uncertainty spread of the models.

Mathematically, the distinguishability statistic $S_x(t)$, as defined by Eq.~\eqref{eq:Sx}, is formally identical to a time-dependent, one-dimensional Mahalanobis distance between the predictive distributions associated with Tv1 and Tv2 solutions. The classical multivariate Mahalanobis distance between two distributions with mean vectors $\boldsymbol{\mu}_1, \boldsymbol{\mu}_2$ and pooled covariance matrix $\mathbf{\Sigma}$ is defined as $D_{\mathrm{M}} = \sqrt{(\boldsymbol{\mu}_1 - \boldsymbol{\mu}_2)^T \mathbf{\Sigma}^{-1} (\boldsymbol{\mu}_1 - \boldsymbol{\mu}_2)}$ \citep{Mahalanobis1936}. For the timing observable $x_{1,\mathrm{eff}}(t)$, the distance between the expected trajectory means $\mu_1(t) = x_{1,\mathrm{eff}}^{\mathrm{Tv1}}(t)$ and $\mu_2(t) = x_{1,\mathrm{eff}}^{\mathrm{Tv2}}(t)$ simplifies directly to the absolute physical separation divided by the square root of the total pooled variance:

\begin{align}
    D_{\mathrm{M}}(t) &= \frac{\left| x_{1,\mathrm{eff}}^{\mathrm{Tv1}}(t) - x_{1,\mathrm{eff}}^{\mathrm{Tv2}}(t) \right|}{\sqrt{\sigma^2_{\mathrm{total}}(t)}} 
    \nonumber \\
    &
    = \frac{\left| x_{1,\mathrm{eff}}^{\mathrm{Tv1}}(t) - x_{1,\mathrm{eff}}^{\mathrm{Tv2}}(t) \right|}{\left[ \sigma^2_{x,\mathrm{Tv1}}(t) + \sigma^2_{x,\mathrm{Tv2}}(t) + \sigma^2_{x,\mathrm{obs}} \right]^{1/2}} \equiv S_x(t)\,.
    \label{eq:mahalanobis_equivalence}
\end{align}

This identification highlights the physical significance of $S_x(t)$ as a scale-invariant metric measured in a variance-weighted uncertainty space. In the absence of full joint posterior samples, we regard ${\cal S}_x(t)$ as a practical forecast of observational distinguishability rather than a formal Bayesian model-selection statistic.

However, there is an additional subtlety in constructing the ${\cal S}_x(t)$ statistic to compare the Tv1 and Tv2 solutions. When we initialize our SA secular equations by directly combining the 1D marginal median parameters, reported in Table~4 of \citetalias{YangHT}, it leads to a geometric inconsistency for solution Tv2. We infer while employing the 1D median values for solution Tv1 ($i_1 = 107.4^\circ$, $i_2 = 147.7^\circ$, $\Delta\Omega = 106.4^\circ$) naturally produce a mutual inclination of $i_{\mathrm{X}} \simeq 83.8^\circ$ that matches the derived value, as reported in \citetalias{YangHT}. However, when we directly plug in the tabulated medians for solution Tv2 ($i_1 = 60.6^\circ$, $i_2 = 29.7^\circ$, $\Delta\Omega = 54.3^\circ$), we obtain a mutual inclination that deviates significantly from the reported centroid of $i_{\mathrm{X}} = 55.0^\circ$ of \citetalias{YangHT}. This apparent inconsistentency  arises due to the fundamental statistical property that non-linear functions of correlated random variables do not commute with the median operator. In other words,  $\mathrm{median}[f(i_1, i_2, \Delta\Omega)] \neq f(\mathrm{median}(i_1), \mathrm{median}(i_2), \mathrm{median}(\Delta\Omega))$. This ensures that the combination of the 1D marginal median parameters directly does not produce a joint posterior sample that preserves the derived mutual inclination $i_{\mathrm{X}}$ for Tv2 as given in \citetalias{YangHT}.

To enforce exact 3D geometric consistency without distorting the mutual inclination, we construct a modified representative configuration, denoted as solution Tv2$^\prime$ from here onward. Retaining the median inclination angles ($i_1 = 60.6^\circ$ and $i_2 = 29.7^\circ$) together with the derived mutual inclination ($i_{\mathrm{X}} = 55.0^\circ$), we recompute the required relative longitude of ascending node $\Delta\Omega^\prime$ via spherical trigonometry, and it leads to
\begin{equation}
    \cos\Delta\Omega^\prime = \frac{\cos i_{\mathrm{X}} - \cos i_1 \cos i_2}{\sin i_1 \sin i_2}\,.
    \label{eq:delta_omega_recomputation}
\end{equation}
Solving Eq.~\eqref{eq:delta_omega_recomputation} yields a representative node longitude difference of $\Delta\Omega^\prime \simeq 70.1^\circ$. This geometrically self-consistent configuration is referred to as solution Tv2$^\prime$ to avoid misrepresenting $\Delta\Omega^\prime$ as a direct parameter fit from \citetalias{YangHT}, ensuring that our forward secular integrations for the $i_{\mathrm{X}} \approx 55^\circ$ branch remain physically and geometrically rigorous.

A few comments are in order before we track the time evolution of the separation statistic defined in Eq.~\eqref{eq:Sx} for the Tv1 and Tv2$^\prime$ solutions. In doing so, we found that particular care is required when translating the parameters inferred from local phenomenological fits into initial conditions for the subsequent secular evolution of the inner binary within our SA framework. This is influenced by the fact that local \texttt{ELL1H} timing over the $\sim$4.6-year FAST baseline yields a precise measurement of the projected semi-major axis, $x_1$, with single-measurement timing uncertainties of $\sigma_{x_1, \rm obs} \approx 0.08\,\mu\mathrm{s}$ as evident in Table~1 of \citetalias{YangHT}. However, when \citetalias{YangHT} perform their HT system fits, listed in their Table~4, the derived median values of $x_1$ at the reference epoch $t_0 = \mathrm{MJD}~59997.5$ exhibit slight variations between solution Tv1 ($x_1 \approx 7.9781319\,\mathrm{lt\text{-}s}$) and solution Tv2 ($x_1 \approx 7.9780683\,\mathrm{lt\text{-}s}$). Therefore, it is important to prevent these static initial median discrepancies from artificially inflating our model distinguishability statistic $S_x(t)$. This prompted us to define the effective timing observable relative to its value at the reference epoch, namely $\Delta x_{1,\mathrm{eff}}(t) \equiv x_{1,\mathrm{eff}}(t) - x_{1,\mathrm{eff}}(t_0)$. Enforcing $\Delta x_{1,\mathrm{eff}}(t_0) = 0$ ensures that initial calibration offsets are accounted for which guarantees that $S_x(t)$ isolates purely the time-accumulated, SA secular dynamical divergence. In contrast, a similar subtraction is neither necessary nor physically permissible for the Laplace-Lagrange parameters, namely $\epsilon_1 = e_1 \sin\omega_1$ and $\epsilon_2 = e_1 \cos\omega_1$. In our vector formalism, given by Eqs.~\eqref{eqs:djde},  $(\epsilon_1, \epsilon_2)$ construct the components of the initial eccentricity vector $\mathbf{e}_1(t_0) = (\epsilon_2,\, \epsilon_1 \cos i_1,\, \epsilon_1 \sin i_1)$ and enter non-linearly into both the coupled secular ODEs $(\dot{\mathbf{e}}_1, \dot{\mathbf{j}}_1)$. Subtracting constant offsets from $\epsilon_1$ or $\epsilon_2$ would violate state-vector orthonormality ($\mathbf{e}_1 \cdot \mathbf{j}_1 = 0$, $e_1^2 + j_1^2 = 1$) and distort the physical shape of the inner orbit. Furthermore, their initial values in Table~4 of \citetalias{YangHT} are virtually identical between Tv1 and Tv2 making an initial subtraction unnecessary.

\begin{figure}
    \centering
    \includegraphics[width=\columnwidth]{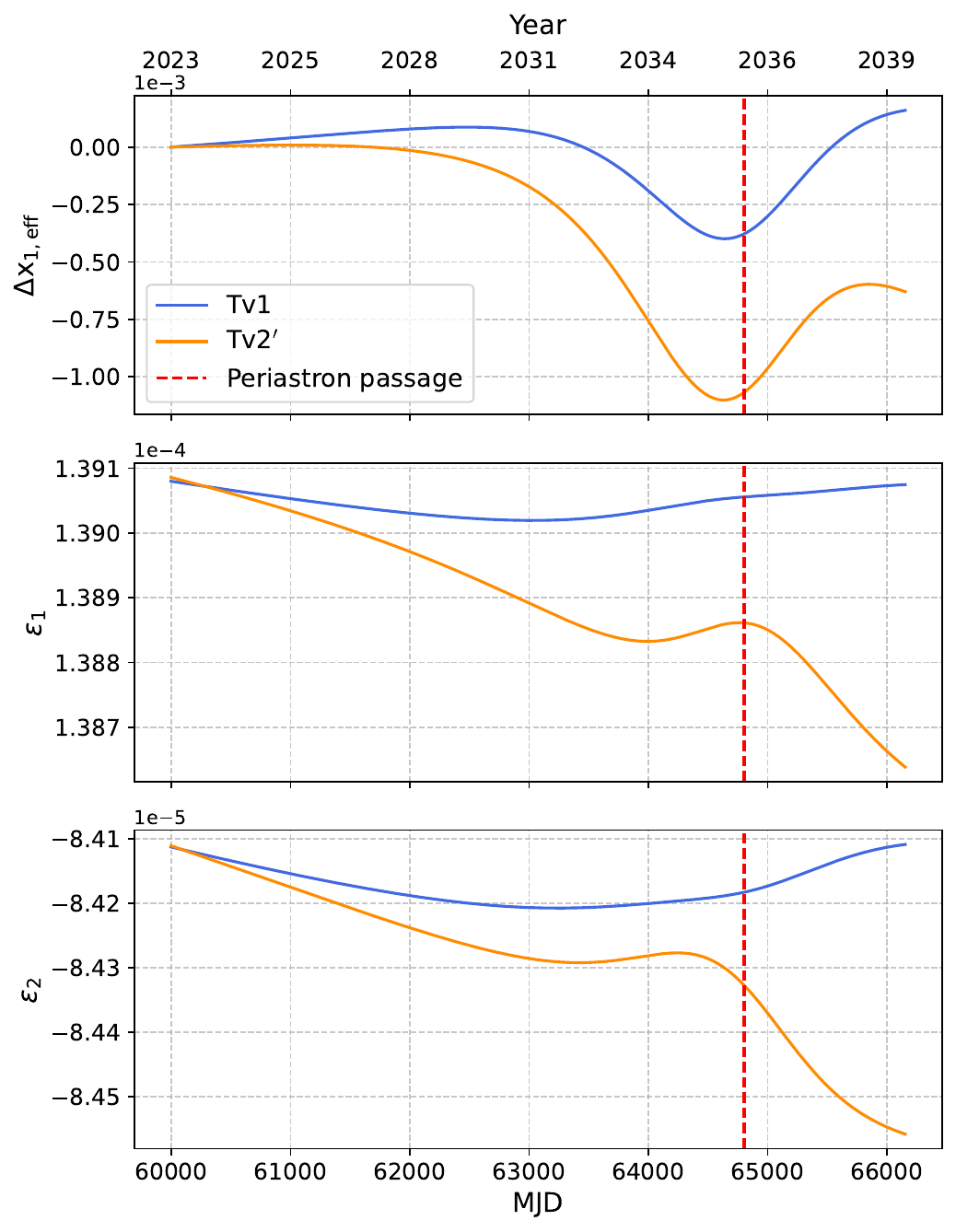}
    \caption{Time evolution of the effective projected semi-major axis and
    the Laplace--Lagrange parameters for the Tv1 and
    Tv2$^\prime$ solutions. The top panel shows
    $\Delta x_{1,\mathrm{eff}}(t)
    = x_{1,\mathrm{eff}}(t)-x_{1,\mathrm{eff}}(t_0)$,
    with the initial value subtracted separately from each solution in order to show the true secular divergence.
    The middle and bottom panels show the corresponding evolution of
    $\epsilon_1=e_1\sin\omega_1$ and
    $\epsilon_2=e_1\cos\omega_1$, respectively. The red dashed line
    marks the outer periastron passage. The increasingly divergent trajectories of the two configurations, particularly near the outer periastron, reflect their distinct secular dynamical evolution across all three observables. 
    }
    \label{fig:tv1_vs_tv2prime}
\end{figure}

We illustrate in Fig.~\ref{fig:tv1_vs_tv2prime} the non-linear variation of $x_{1,\mathrm{eff}}(t)$ after subtracting $x_{1,\mathrm{eff}}(t_0)$ for solutions Tv1 and Tv2$^\prime$. The visibile dramatic bifurcation between the two plots as our HT  approaches its wide outer periastron stems from two coupled physical mechanisms: nodal precession and outer kinematic velocity modulations. First, because solutions Tv1 ($i_{\mathrm{X}} \simeq 83.8^\circ$) and Tv2$^\prime$ ($i_{\mathrm{X}} \simeq 55.0^\circ$) represent fundamentally different inter binary mutual inclinations, the quadrupolar secular potential $\Phi_{\mathrm{SA}} \propto R^{-3}(t)$ exerts distinct tidal torques on the inner orbital plane. This drives markedly different rates of inclination evolution $\dot{i}_1(t)$, which translate into divergent geometric projection factors $x_{1,\mathrm{SA}}(t) \propto \sin i_1(t)$ via the nodal derivative $\dot{x}_{1,\mathrm{SA}} = x_{1,\mathrm{SA}} \cot i_1 \dot{i}_1$. Second, the observable effective projected semi-major axis incorporates classical Doppler and transverse-Doppler corrections, as noted in Eq.~\eqref{eq:xeff}, tracking the inner binary barycenter's motion in the wide orbit. Because Tv1 ($i_2 = 147.7^\circ$) and Tv2$^\prime$ ($i_2 = 29.7^\circ$) possess different outer inclinations and mass scales ($M_{\star}, M_{\mathrm{in}}$), the outer velocity projections $(v_{2,\parallel}, v_{2,y})$ diverge strongly, amplifying the physical trajectory separation $\Delta x_{1,\mathrm{eff}}(t)$ to $ \sim103\,\mu\mathrm{s}$ even by late 2028.

\begin{figure}
    \centering
    \includegraphics[width=\columnwidth]{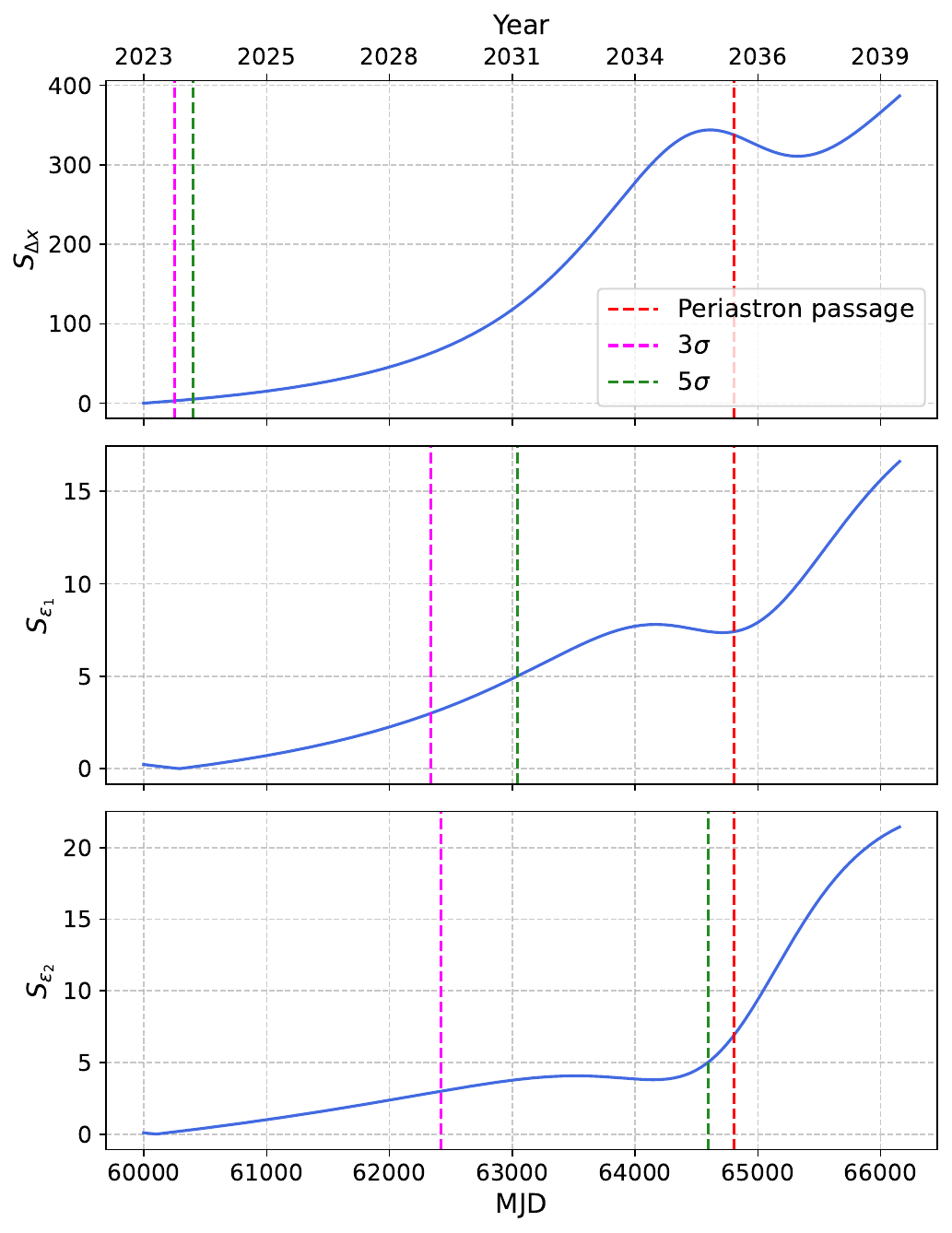}
    \caption{Time evolution of the separation statistics, as defined by Eq.~\eqref{eq:Sx}, between the Tv1 and Tv2$^\prime$ timing solutions for the inner-binary dynamical variables: $\Delta x_{1,\rm eff} \equiv x_{1,\rm eff} (t) - x_{1,\rm eff}(t_0)$, $\epsilon_1 (t)$, and $\epsilon_2 (t)$. The panels show the corresponding separation statistics
    $S_{\Delta x}$, $S_{\epsilon_1}$, and $S_{\epsilon_2}$, respectively, as a
    function of time. The vertical dashed lines mark the epochs at which the predicted solutions become distinguishable at the $3\sigma$ (pink) and $5\sigma$ (green) levels, while the red dashed line indicates the predicted periastron passage of the outer orbit. Notably, the $S_{\Delta x}$ statistic in the top panel reaches the $5\sigma$-level within the current FAST timing baseline, which spans approximately $4.6$~yr, indicating that the predicted evolution of $x_{1,\rm eff}$ between the two configurations should already be distinguishable at high significance using the currently available timing data. The increasing separation between the two configurations toward the outer periastron demonstrates the growing
    potential of future timing observations to break the present geometrical degeneracy.}
    \label{fig:separation_statistics}
\end{figure}

With the zero-point calibration established to isolate the true secular divergence, we are now in a position to evaluate the forward model distinguishability. In Fig.~\ref{fig:separation_statistics}, we track the time evolution of the model-penalized separation statistics $S(t)$, as defined in Eq.~\eqref{eq:Sx}, for solutions Tv1 and Tv2$^\prime$ \citepalias{YangHT}. Among the dynamical observables considered, the calibrated effective projected semi-major axis of the inner binary, $\Delta x_{1, \mathrm{eff}}(t)$, emerges as the single most powerful discriminant between the two proposed orbital geometries. As noted earlier, this behaviour is driven by the coupled effects of three-dimensional nodal precession, through $\dot{i}_1$, and outer-orbit kinematic velocity modulations, through $v_{2,\parallel}$ and $v_{2,y}$, which become particularly pronounced near periastron. Together, these effect cause $\Delta x_{1,\mathrm{eff}}(t)$ to accumulate rapidly, producing a decisive separation between the competing trajectories that soon exceeds the single-measurement precision achievable with FAST. Importantly, our estimates indicate that the existing FAST timing baseline may already contain sufficient information to distinguish between these solutions, particularly if the ToAs are reanalyzed using the SA prescription developed here. Such an analysis is potentially advantageous because the SA framework predicts the orbital evolution directly from the underlying three-body dynamics, rather than representing it through additional phenomenological parameters, such as successive higher-order derivatives of the projected semi-major axis and inner orbital frequency, which can introduce strong parameter covariances.

Remarkably, the predicted separation of all four observables becomes significant before the expected periastron passage of the subgiant tertiary. The earlier crossing of the adopted significance threshold by $\epsilon_1$ relative to $\epsilon_2$ can be understood from the substantially different magnitudes of their predicted dynamical variations relative to their measurement uncertainties. These forecasts provide a strong motivation for continued high-precision FAST timing of this unique system, particularly during its approach to outer periastron. In the following sections, we extend this dynamical picture beyond pulsar timing and explore the complementary multi-band observational opportunities offered by the system.

\section{Probing Multi-Technique Implications}
\label{sec:multi-technique}

The presence of the optically visible tertiary, Gaia~DR3~171833395176138240 ($G\simeq16.65$~mag) (\citetalias{YangHT}), makes the HT associated with PSR~J0435$+$3233 a particularly compelling multi-band system, accessible through both precision radio/$\gamma$-ray timing and optical spectroscopic and astrometric observations. A possible spectroscopic monitoring of the optically identified tertiary should provide an independent and complementary dynamical view of the system. Measurements of the stellar radial-velocity (RV) curve trace the opposite side of the triple's barycentric motion and determine the tertiary's RV amplitude, $K_{\star}$. \textit{Gaia} astrometry supplies the third and crucial component by tracing the motion of the tertiary on the plane of the sky, thereby providing direct constraints on the orientation and angular scale of the outer orbit. In what follows, we demonstrate how our SA formulation naturally connects these complementary observational probes by mapping the same underlying three-body dynamics onto the timing, spectroscopic, and astrometric signatures of this unique HT.

\subsection{Probing Radial Velocity Implications }
\label{sec:LAMOST_HT}

We have shown that combining our SA framework with precision FAST timing provides a detailed description of the motion of the inner binary. In this respect, precision pulsar timing with FAST may be viewed as the dynamical analogue of single-lined spectroscopy: the pulse ToAs provide an exquisitely precise tracer of the line-of-sight motion of the pulsar, and hence of the inner-binary barycenter along the outer orbit. When interpreted through our SA framework, these measurements essentially provide the orbital motion and corresponding RV amplitude of the inner-binary barycenter. The presence of the optically visible tertiary, Gaia~DR3~171833395176138240 ($G\simeq16.65$~mag), offers the complementary ``second line'' as the high-resolution spectroscopic monitoring can directly measure the tertiary's RV curve and its amplitude, $K_\star$. Therefore, the combination is rather analogous to a double-lined spectroscopic system, with pulsar timing and optical spectroscopy tracing the two sides of the wide outer orbit about their common barycenter.

Recall that in our SA secular approach $\mathbf{R}(t)$ stands for the 3D relative separation vector pointing from the inner-binary barycenter to the tertiary star, and its time derivative  $\mathbf{V}(t) = \dot{\mathbf{R}}(t)$ denote the outer relative velocity vector. Mass-weighting $\mathbf{V}(t)$ and projecting along the line-of-sight unit vector $\hat{\mathbf z}$ leads to 
\begin{equation}
    v_{\star,\parallel}(t) = \gamma_{\mathrm{sys}} + \frac{M_{\mathrm{in}}}{M_{\mathrm{in}} + M_{\mathrm{\star}}} \left( \mathbf{V}(t) \cdot \hat{\mathbf z} \right)\,,
    \label{eq:vstar_vector}
\end{equation}
where $M_{\mathrm{in}}$ is the total mass of the inner compact binary and $M_{\star}$ is the tertiary star's mass. Further, $\gamma_{\mathrm{sys}}$ is the systemic radial velocity of the triple system's barycenter relative to the Solar System barycenter.

We now express the above expression  in terms of the outer Keplerian orbital elements, and the tertiary line-of-sight velocity curve takes the classical spectroscopic form \citep{MurrayCorreia2010}:
\begin{equation}
    v_{\star,\parallel}(t) = \gamma_{\mathrm{sys}} + K_{\star} \left[ \cos\left(\omega_2 + f_2\right) + e_2 \cos\omega_2 \right]\,,
    \label{eq:vstar_keplerian}
\end{equation}
where $K_{\star}$ is the semi-amplitude of the velocity of the tertiary companion around the barycenter of the triple-system. For our HT system, it is given by \citep{MurrayCorreia2010} 
\begin{equation}
    K_{\mathrm{\star}} = \frac{M_{\mathrm{in}}}{M_{\mathrm{in}} + M_{\star}} \frac{2\pi a_{\mathrm{out}} \sin i_2}{P_2 \sqrt{1 - e_2^2}}\,.
    \label{eq:kstar}
\end{equation}

This expression prompted us to explore whether optical spectroscopy can independently resolve the geometrical ambiguity of PSR~J0435$+$3233, as the two configurations possess different outer orbital inclination angle, and distinct component masses.

\begin{figure*}
    \centering
    \includegraphics[width=0.90\textwidth]{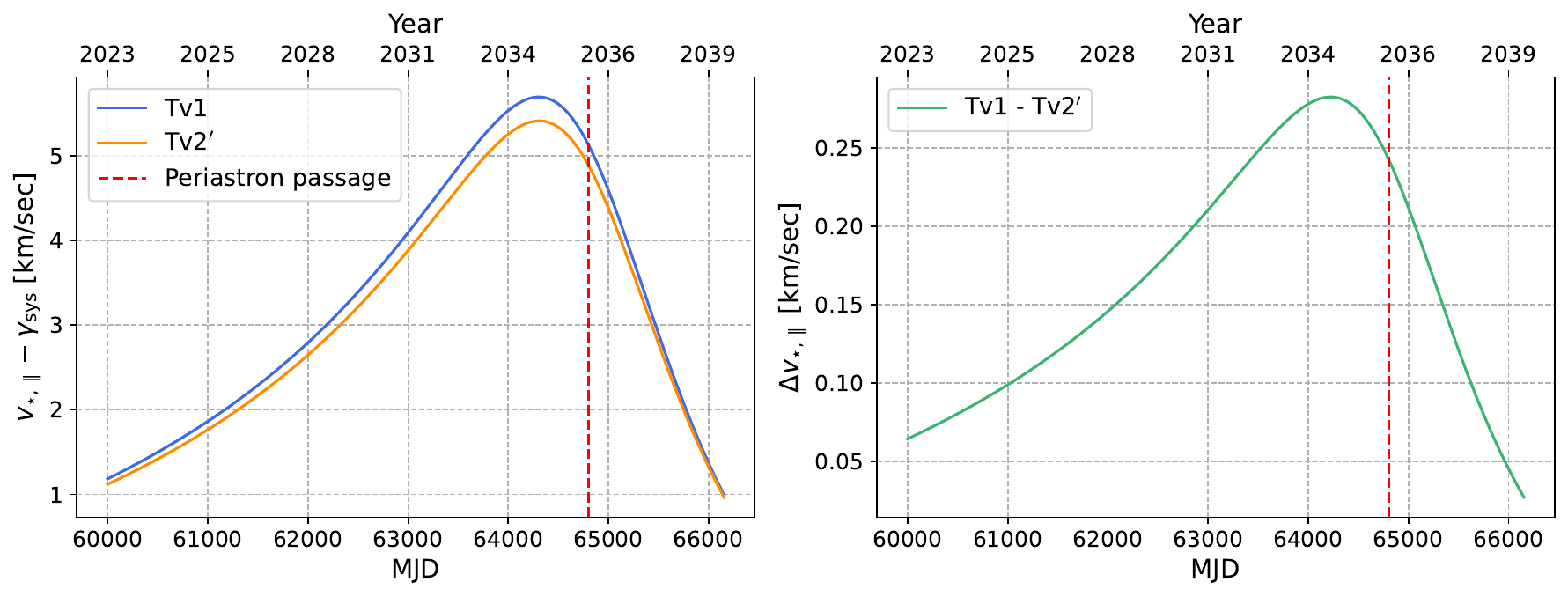}
    \caption{Predicted line-of-sight velocity of the tertiary star for the Tv1 and Tv2$^\prime$ configurations, obtained by propagating the outer orbit using the timing solutions listed in Table~4 of \citetalias{YangHT}. The left panel shows the predicted barycentric line-of-sight velocity of the tertiary star, with the two configurations represented by the blue and orange curves, respectively. The right panel shows the corresponding difference $\Delta v_{\star,\parallel} \equiv v_{\star,\parallel}^{\rm Tv1}-v_{\star,\parallel}^{\rm Tv2^\prime}$, which quantifies the predicted spectroscopic separation between the two configurations. The red dashed line marks the predicted periastron passage of the outer orbit. The difference between the two predictions reaches $\sim0.3$~km~s$^{-1}$ within the chosen time interval, indicating that high-precision spectroscopic monitoring of the tertiary could provide an independent constraint on the present geometrical degeneracy.
    }
    \label{fig:star_radial_velocity}
\end{figure*}

In Fig.~\ref{fig:star_radial_velocity}, we show the predicted line-of-sight radial-velocity trajectories of the tertiary, $v_{\star,\parallel}(t)$, obtained from Eq.~\eqref{eq:vstar_keplerian} for the Tv1 and Tv2$^\prime$ solutions, after subtracting $\gamma_{\rm sys}$. Although the two solutions predict appreciable secular evolution of the tertiary's radial velocity, their differential signature,
\begin{equation}
\Delta v_{\star,\parallel}(t) \equiv
v_{\star,\parallel}^{\rm Tv1}(t)-
v_{\star,\parallel}^{\rm Tv2^\prime}(t)\,,
\label{eq:delta_vstar}
\end{equation}
remains rather small, ranging from approximately $0.06$ to $\sim 0.3\,\mathrm{km\,s^{-1}}$ over the coming decade. Therefore, a useful independent route for discriminating between the two FAST timing solutions would be high-resolution optical spectroscopy of the G-type tertiary.

We note that, although Tv1 and Tv2$^\prime$ have very similar outer orbital periods and eccentricities, they differ appreciably in their component masses and three-dimensional orbital geometries. These differences lead to distinct, albeit closely spaced, radial-velocity trajectories for the tertiary. Using the outer-orbit parameters inferred for the two a, we obtain tertiary radial-velocity semi-amplitudes of approximately $K_\star\simeq3.78$ and $3.59~{\rm km\,s^{-1}}$ for Tv1 and Tv2$^\prime$, respectively. This is why Eq.~\eqref{eq:delta_vstar} predicts that the instantaneous line-of-sight velocities of the two solutions can differ by as much as $\Delta v_{\star,\parallel}\sim0.3~{\rm km\,s^{-1}}$ over the relevant observing interval. A high-resolution spectrum obtained in the near future would therefore establish a valuable absolute-RV reference epoch, while repeated measurements over the following years could trace the orbital acceleration and test which of the two timing-predicted trajectories is followed. Such observations require high-resolution, high-stability spectroscopy on a large-aperture telescope, with sub-${\rm km\,s^{-1}}$ precision, and ideally a few $0.1~{\rm km\,s^{-1}}$ per epoch, to exploit the predicted separation between Tv1 and Tv2$^\prime$.

Although the expected radial-velocity precision of LAMOST is unlikely to be sufficient to discriminate directly between the closely spaced Tv1 and Tv2$^\prime$ trajectories, continued spectroscopic monitoring of the tertiary would nevertheless provide valuable complementary information. The outer orbit spans approximately 74~yr and is highly eccentric, so multi-epoch spectroscopy over a sufficiently long baseline could help constrain the systemic velocity, $\gamma_{\rm sys}$, and trace the long-term Keplerian radial-velocity evolution of the tertiary as it approaches the $\sim$2036 periastron passage. Such measurements would provide an independent spectroscopic constraint on the outer orbit and its orbital phase, test the overall RV trend predicted by the timing-derived solutions. Such an effort will provide a long temporal baseline that can be complemented by occasional high-resolution spectroscopy capable of resolving the much smaller Tv1-Tv2$^\prime$ radial velocity separation. When combined with FAST timing, these measurements would contribute to a self-consistent dynamical description of the HT in PSR~J0435$+$3233. A joint Bayesian timing--spectroscopic analysis would therefore involve 17 parameters, with $\gamma_{\rm sys}$ entering as an additional parameter alongside the 16-dimensional dynamical parameter set defined in Eq.~\eqref{eq:mcmc_orbital_vector}.
 
We note in passing that optical spectroscopy has previously played a decisive role in establishing the nature of another unusual MSP system. Spectroscopic observations of the companion to PSR~J1903$+$0327, obtained with the Gemini Multi-Object Spectrograph on Gemini-North \citep{Gemini}, enabled one of us and collaborators to demonstrate that the system is not a presently bound HT, but instead consists of a Galactic MSP in a highly eccentric orbit ($e\simeq0.44$) with a main-sequence companion \citep{DC1903,Gopakumar2009,Juthika1903}.

As noted earlier, precision FAST timing of PSR~J0435$+$3233 and optical spectroscopy of the G-type tertiary provide complementary probes of the wide orbit, making the system dynamically analogous to a double-lined spectroscopic binary. However,  the motion of the inner-binary barycenter of our HT is constrained indirectly by pulsar timing, through the tertiary-induced line-of-sight acceleration and its higher derivatives, rather than through a direct radial-velocity measurement. Our three-dimensional SA dynamical model then predicts the corresponding semi-amplitude of the radial-velocity of inner binary's barycenter,
\begin{equation}
 K_{2}
 = \frac{M_{\star}}{M_{\mathrm{in}} + M_{\star}}
 \frac{2\pi a_{\rm out}\sin i_2}
 {P_2\sqrt{1-e_2^2}}\,.
 \label{eq:k2}
\end{equation}

This quantity is not an independent parameter of our SA formulation, but is uniquely determined by $M_{\rm PSR}$, $M_{\rm WD}$, $M_{\star}$, and $P_2$, and can therefore be derived self-consistently from our timing-based dynamical solution. In contrast, the optical spectroscopy provides the complementary semi-amplitude, $K_\star$, of the tertiary. Since these two quantities trace opposite sides of the same outer orbit, their ratio directly yields
\begin{equation}
 \frac{K_{2}}{K_\star}
 =\frac{M_\star}{M_{\rm in}}\,.
 \label{eq:mass_ratio}
\end{equation}

Thus, combining the timing-constrained outer motion with spectroscopy of the tertiary provides an independent constraint on the outer mass ratio. This differs from the approach employed for PSR~J1903$+$0327, where pulsar timing and optical radial-velocity measurements of the stellar companion were combined to constrain the properties of a single binary orbit \citep{Juthika1903}.

This should have interesting implications. Recall that $M_{\rm PSR}$, $M_{\rm WD}$, and $M_\star$ enter directly as physical parameters governing the three-body dynamics in our SA approach. The secular perturbations measured through FAST timing, together with the precisely determined inner-binary mass function, therefore constrain different combinations of these three masses. A sufficiently precise spectroscopic measurement of the tertiary radial-velocity amplitude provides an additional, largely independent constraint through the outer-orbit mass ratio. The combination of these timing, and spectroscopic constraints can therefore substantially over-constrain the system, enabling significantly improved determinations of the individual masses $(M_{\rm PSR},M_{\rm WD},M_\star)$ and providing an important internal consistency test of the three-body model. A closely related constraint is provided by the double pulsar, PSR~J0737$-$3039A/B, where timing both pulsars yields the projected semimajor axes of their respective barycentric orbits. Their ratio directly determines the mass ratio, $M_{\rm A}/M_{\rm B}=x_{\rm B}/x_{\rm A}$, to leading post-Keplerian accuracy, providing an exceptionally clean dynamical constraint independent of the usual post-Keplerian parameters
\citep{Burgay2003,Lyne2004,Kramer2006}.

We now explore how existing and forthcoming \textit{Gaia} data releases can provide complementary astrometric constraints on the three-dimensional architecture and dynamical evolution of the HT associated with PSR~J0435$+$3233.

\subsection{Implications for Gaia Data Sets }
\label{sec:GAIA_HT}

In this subsection, we argue that the utility of the SA secular framework extends beyond radio pulsar timing and spectroscopic radial velocities. Note that this approach allows us to provide a mathematical bridge to directly model the companion star's astrometric motion on the sky plane as observed by the \textit{Gaia} mission~(\citetalias{Gaia2016}). This mission provides high-precision astrometric measurements for more than a billion Galactic sources in the currently available \textit{Gaia} Data Release 3~(DR3; \citetalias{GaiaDR32022}). We note that this Data Release allowed \citetalias{YangHT} to identify the optically visible tertiary associated with PSR~J0435$+$3233 as Gaia~DR3~171833395176138240. The forthcoming \textit{Gaia} DR4, expected in late 2026, should provide substantially improved astrometric information because of its longer observational baseline. We now examine how existing and forthcoming data releases can be incorporated into our SA framework to track the sky-plane motion of the optically visible tertiary and thereby provide independent constraints on the three-dimensional geometry of the HT associated with PSR~J0435$+$3233.

At any instant $t$, our SA secular framework computes the relative position vector $\mathbf{R}(t)$ of the tertiary star relative to the inner binary's barycenter.
The three-dimensional barycentric position vector of the companion star relative to the triple system barycenter is given by the usual mass-weighted relation as 
\begin{equation}
    \mathbf{r}_{\mathrm{\star}}(t) = \frac{M_{\mathrm{in}}}{M_{\mathrm{in}} + M_{\mathrm{\star}}} \mathbf{R}(t)\,.
    \label{eq:rstar_tripbary}
\end{equation}

By projecting $\mathbf{r}_{\mathrm{\star}}(t)$ onto the plane of the sky, the SA model directly predicts the time-dependent angular offsets in Right Ascension ($\Delta \alpha^* \equiv \Delta \alpha \cos \delta$) and Declination ($\Delta \delta$) relative to the system's barycentric proper motion:
\begin{subequations}
\label{eq:astrometric_offsets}
\begin{align}
    \Delta \alpha^*(t)
    &= \frac{\mathbf{r}_{\mathrm{\star}}(t) \cdot \hat{\mathbf{x}}_{\mathrm{sky}}}{D}
    = \frac{M_{\mathrm{in}}}{D(M_{\mathrm{in}} + M_{\mathrm{\star}})}
    \left( \mathbf{R}(t) \cdot \hat{\mathbf{x}}_{\mathrm{sky}} \right)\,,
    \label{eq:delta_alpha}
    \\
    \Delta \delta(t)
    &= \frac{\mathbf{r}_{\mathrm{\star}}(t) \cdot \hat{\mathbf{y}}_{\mathrm{sky}}}{D}
    = \frac{M_{\mathrm{in}}}{D(M_{\mathrm{in}} + M_{\mathrm{\star}})}
    \left( \mathbf{R}(t) \cdot \hat{\mathbf{y}}_{\mathrm{sky}} \right)\,,
    \label{eq:delta_delta}
\end{align}
\end{subequations}
where $D \approx 2.1 \pm 0.4$\,kpc (\citetalias{YangHT}) is the parallax distance to the system, and $\hat{\mathbf{x}}_{\mathrm{sky}}$ and $\hat{\mathbf{y}}_{\mathrm{sky}}$ are the East and North unit vectors in the sky plane. We note in passing that the conversion of the 3D relative orbital unit vector $\hat{\mathbf{R}}$ from the outer orbital plane to the celestial sky-plane frame is carried out via standard Euler rotations through the outer longitude of the ascending node ($\Omega_2$), the outer orbital inclination ($i_2$), and the outer argument of the periastron ($\omega_2$) \citep{DS1988}.

Because the companion's outer orbital period is extremely long ($P_2 \approx 74$\,years) relative to \textit{Gaia}'s active observation baselines, a standard closed Keplerian orbit cannot yet be fully resolved from \textit{Gaia}'s astrometric data alone. Instead, in the \textit{Gaia} DR3 Non-Single Star (NSS) catalog~(\citealt{GaiaDR3_astrometric}, \citetalias{GaiaDR32022}), such long-period binary and multiple star candidates are modeled using polynomial proper motion derivatives. These are hosted in the \texttt{gaiadr3.nss\_acceleration\_astro} table as acceleration (7-parameter, quadratic in time) or jerk (9-parameter, cubic in time) solutions \citep{GaiaDR3_astrometric}:
\begin{equation}
    \boldsymbol{\mu}(t) = \boldsymbol{\mu}_0 + \dot{\boldsymbol \mu} (t - t_0) + \frac{1}{2} \ddot{\boldsymbol \mu} (t - t_0)^2 + \dots \,.
    \label{eq:gaiadr3_astro}
\end{equation}
These polynomial coefficients ($\dot{\boldsymbol \mu}$ and $\ddot{\boldsymbol \mu}$) are purely phenomenological. The SA secular formalism allows these abstract proper motion derivatives to be translated directly into fundamental physical coordinates of the triple system. By taking successive time derivatives of our projected position equations, we establish the direct physical mapping:
\begin{subequations}
\label{eq:mu_derivatives}
\begin{align}
    \dot{\boldsymbol{\mu}}
    &\equiv \frac{\ddot{\mathbf{r}}_{\mathrm{\star},\perp}(t_0)}{D}
    = \frac{M_{\mathrm{in}}}{D(M_{\mathrm{in}} + M_{\star})}
    \ddot{\mathbf{R}}_{\perp}(t_0)\,,
    \label{eq:mu_dot}
    \\
    \ddot{\boldsymbol{\mu}}
    &\equiv \frac{\dddot{\mathbf{r}}_{\mathrm{\star},\perp}(t_0)}{D}
    = \frac{M_{\mathrm{in}}}{D(M_{\mathrm{in}} + M_{\star})}
    \dddot{\mathbf{R}}_{\perp}(t_0)\,,
    \label{eq:mu_ddot}
\end{align}
\end{subequations}
where the subscript $\perp$ denotes projection onto the sky plane. This mapping allows \textit{Gaia}'s measured astrometric acceleration and jerk to directly constrain the mass ratio and outer orbit geometry in a physically consistent manner.

There is another important implication for the role of \textit{Gaia} astrometry. Both FAST timing and optical spectroscopy primarily probe line-of-sight projections of the orbital motion and therefore retain an inherent reflection degeneracy with respect to the plane of the sky, $i\leftrightarrow180^\circ-i$. In particular, the orthometric Shapiro parameter, $\varsigma=\sin i_1/(1+|\cos i_1|)$, and the spectroscopic radial-velocity semi-amplitude, $K_\star\propto\sin i_2$, are insensitive to the sign of $\cos i_1$ and $\cos i_2$, respectively. Consequently, these line-of-sight measurements alone cannot distinguish between the corresponding mirror orbital configurations, leaving the sense of the three-dimensional orbital orientation ambiguous. Astrometric measurements of the sky-plane motion with \textit{Gaia} provide a complementary means of lifting this degeneracy.

High-precision \textit{Gaia} astrometry provides the crucial sky-plane information needed to break this reflection degeneracy. For the G-type subgiant tertiary, namely Gaia~DR3~171833395176138240, \textit{Gaia} measures its position, parallax ($\varpi=0.466\pm0.078\,\mathrm{mas}$), and two-dimensional proper motion $(\mu_{\alpha^\ast},\mu_\delta)$ on the celestial sphere, while sufficiently precise astrometric monitoring can additionally constrain the sky-plane curvature associated with its outer orbital motion. A reflection of the orbital plane through the plane of the sky, $i_2\rightarrow180^\circ-i_2$, reverses the sign of the line-of-sight component of the orbital angular momentum, $j_{2,z}=|\mathbf{j}_2|\cos i_2$, while leaving $\sin i_2$ unchanged. Consequently, the combination of the astrometrically determined sky-plane orbit and its direction of motion with the line-of-sight information supplied by spectroscopy can distinguish the two mirror configurations and determine the sign of $\cos i_2$, thereby fixing the three-dimensional orientation of the outer orbit. This implies that \textit{Gaia} astrometry supplemented with spectroscopy can directly resolve the mirror ambiguity of the outer orbit. 

\begin{figure}
    \centering \includegraphics[width=\columnwidth]{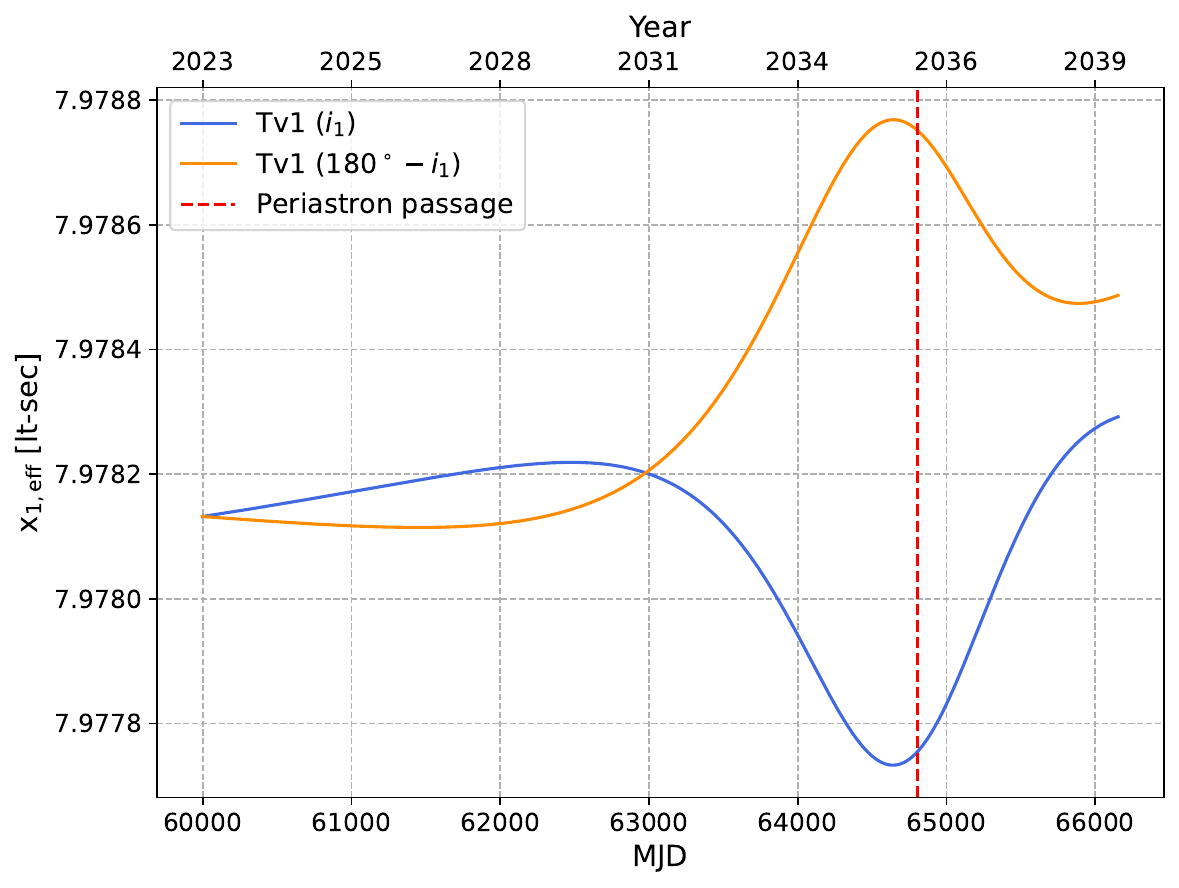}
    \caption{
        Evolution of the effective projected semi-major axis, $x_{1,\rm eff}$, for the two mirror-degenerate inner-orbit configurations, corresponding to $i_1$ and $180^\circ-i_1$, respectively. The three-dimensional configuration of the outer orbit is held fixed to that of the Tv1 solution, motivated by the prospect that the outer-orbit geometry can eventually be independently constrained through complementary multi-technique observations of the G-type star. The two inner-orbit configurations are then evolved independently, while retaining this common outer-orbit geometry, using the same SA three-body equations. Although the two configurations are indistinguishable at the reference epoch through binary timing delays, such as the R\"omer and Shapiro delays, whose relevant inclination dependence involves $\sin i_1$, their subsequent $x_{1,\rm eff}(t)$ trajectories diverge appreciably. This arises because the two mirror configurations of the inner binary correspond to substantially different mutual inclinations, $i_{\rm X}$, and therefore experience distinct three-body perturbations under their subsequent SA evolution. The dashed vertical line marks the outer-orbit periastron passage.     
    }
    \label{fig:inner_mirror_xeff}
\end{figure}

We now investigate whether the inner-orbit mirror ambiguity can be resolved through SA  three-body evolution. This is done by pursuing an additional SA experiment in which the three-dimensional configuration of the outer orbit is held fixed, as would be possible once its orientation is constrained astrometrically. We then evolve two inner-orbit configurations related by the reflection $i_1\rightarrow 180^\circ -i_1$, which are indistinguishable through observables depending only on $\sin i_1$ at the reference epoch. This numerical experiment is motivated by the explicitly orientation-dependent structure of the SA equations, which contain terms such as $({\mathbf j}_1\!\cdot\!\hat{\mathbf R}) ({\mathbf j}_1\!\times\!\hat{\mathbf R})$. Under inner-orbit reflection $i_1\rightarrow 180^\circ -i_1$, while the three-dimensional outer orbit is kept fixed, these combinations are not, in general, invariant. This suggests that the tertiary-induced torque may drive measurably different subsequent evolution of the two initially mirror-degenerate inner configurations. Influenced by these considerations, we  propagate both inner binary configurations using the same SA equations and compare their predicted $x_{1, \rm eff}(t)$ trajectories. As shown in Fig.~\ref{fig:inner_mirror_xeff}, the two initially mirror-degenerate configurations follow substantially different $x_{1,\rm eff}(t)$ trajectories when evolved under the same SA equations. The difference between the two configurations becomes particularly pronounced around the outer periastron passage, where the tertiary-induced interaction is strongest. This demonstrates that, once the three-dimensional orientation of the outer orbit is sufficiently constrained through the combination of FAST timing, LAMOST spectroscopy, and \textit{Gaia} astrometry, the subsequent time-dependent three-body dynamics can break the remaining inner-orbit mirror degeneracy. In particular, the two mirror configurations predict distinct evolutions of $x_{1,\rm eff}(t)$, which can then be distinguished using FAST timing alone.

We conclude by emphasizing that our SA framework provides a natural means of combining the sub-microsecond timing precision of FAST, the potential for long-term RV monitoring with LAMOST, and the high-precision astrometric and geometrical constraints provided by \textit{Gaia}. Together, these complementary observational probes, interpreted within a common three-body dynamical framework, have the potential to transform PSR~J0435$+$3233 into an exceptionally well-constrained laboratory for investigating hierarchical three-body dynamics, compact-binary and stellar evolution, and gravitational physics.

\section{Discussion and Conclusions}
\label{sec:discussion}

We have presented a physically self-consistent SA framework for the HT hosting the FAST-discovered PSR~J0435$+$3233 (\citealt{Wu2026}; \citetalias{YangHT}). By averaging over the rapid $\simeq8$-day inner orbit while retaining the instantaneous three-dimensional position of the G-type subgiant along its eccentric $\simeq74$-yr outer orbit, our formulation captures the explicit time-dependent perturbation exerted by the tertiary. The resulting evolution is obtained through direct numerical integration of the coupled Milankovitch equations \citep{Milankovitch1939} for $(\mathbf{e}_1,\mathbf{j}_1)$, supplemented by the leading 1PN relativistic periastron precession, influenced by \cite{Luo2016, Tremaine2009,Damour1992}. This allows one to replace a phenomenological expansion in successive orbital derivatives with an evolution governed directly by the underlying three-body dynamics. For implementation within a phase-coherent timing model, particularly during the approach to the 2036 outer periastron, we additionally incorporate the outer-orbit kinematic corrections and the non-periodic $O(e_1)$ contribution to the \texttt{ELL1k} R{\o}mer delay, $-(3/2)(x_{1, \rm eff}/c)\epsilon_1(t)$
\citep{Susobhanan2018}.

Using the time-dependent distinguishability statistic ${\cal S}_x(t)$, we find that the effective projected semi-major axis, $x_{1, \rm eff}(t)$, provides the most sensitive timing observable for discriminating between the presently viable Tv1 and Tv2 configurations identified by \citetalias{YangHT}. Notably, our SA framework reveals that, although the Tv1 and Tv2 solutions remain nearly degenerate over the existing $\sim$4.6-yr FAST timing baseline, their predicted trajectories should become distinguishable well before the 2036 outer periastron. Interestingly, our forecasts indicate that the existing $\sim$4.6-yr FAST timing baseline may already contain sufficient information to distinguish these solutions at high statistical significance when the pulse ToAs are reanalyzed within the SA-informed timing framework. 

Additionally, we demonstrate that the same three-dimensional dynamical framework  provides a natural connection between FAST timing,  optical spectroscopy of the tertiary, and \textit{Gaia} astrometry. While timing and radial-velocity measurements predominantly probe line-of-sight orbital motion, astrometric measurements supply the complementary sky-plane information needed to constrain the absolute orbital orientation and address the associated $i\leftrightarrow180^\circ-i$ reflection degeneracies. Therefore, a joint dynamical analysis of these complementary datasets by employing our SA formulation should substantially tighten the three-dimensional geometry and component masses while providing stringent internal consistency tests
of the HT solution. 

A particularly compelling application of PSR~J0435$+$3233 is a strong-field test of the universality of free fall (UFF), following the classic proposal of \citet{DS1991}. If strongly self-gravitating bodies fall differently in an external gravitational field, the neutron star and its WD companion experience a differential acceleration that polarizes the inner orbit, generating a forced component of its eccentricity vector \citep{DS1991}. We note that the coplanar HT in  PSR~J0337$+$1715 provides a particularly clean realization of this strong-field Nordtvedt experiment, with the outer WD supplying the external gravitational field \citep{KenN1968,Archibald2018}. No evidence for such a differential acceleration has been found. Most recently, \citet{Voisin2025} constrained the UFF-violation parameter to $|\Delta|<1.5\times10^{-6}$ or $|\Delta|<2.3\times10^{-6}$ at 95\% confidence, depending on whether the observed long-period timing signal is modeled as a circum-triple planet or as achromatic red noise. This model dependence highlights the importance of an accurate dynamical description when searching for such subtle strong-field signatures.

PSR~J0435$+$3233 may eventually provide an even more sensitive realization of this experiment \citep{Freire2026}. Its inner orbit is timed with FAST at high precision; its projected pulsar semi-major axis is substantially larger than that of PSR~J0337$+$1715, and its optically visible tertiary provides independent spectroscopic and astrometric probes of the outer orbit. More importantly, its long and eccentric outer orbit causes the external gravitational field to vary predictably in both magnitude and direction, providing a time-dependent UFF signature rather than a purely static orbital polarization. Our SA formulation is particularly well suited to exploiting this feature because it directly evolves the eccentricity vector $\mathbf{e}_1(t)$ while retaining the instantaneous position of the tertiary. A UFF-violating differential acceleration can therefore be incorporated as an additional forcing term and constrained simultaneously with the standard Newtonian three-body and relativistic contributions. Continued FAST timing, particularly toward the 2036 outer periastron, may therefore allow PSR~J0435$+$3233 to improve upon existing strong-field UF constraints.

PSR~J0435$+$3233 also provides an unusually well-constrained laboratory for studying compact-binary and stellar evolution. The $\simeq8$-day MSP--helium-WD inner binary carries the imprint of the mass-transfer episode that recycled the pulsar, and its present orbital period and WD mass can be confronted directly with the expected $P_{\rm orb}$--$M_{\rm WD}$ relation for recycled pulsar binaries \citep{TaurisS1999,ITL2014}. The surviving G-type subgiant tertiary provides complementary information about the evolutionary history of the primordial triple, whose present architecture can retain signatures of stellar mass loss, neutron-star formation and its associated natal kick, subsequent mass transfer, and long-term three-body evolution (\citealt{PCZ2012, Toonen2016}; \citetalias{YangHT}).

Measurements of the three component masses, the inner and outer orbital elements, and particularly the mutual orbital inclination can therefore be confronted with evolutionary calculations aimed at reconstructing the formation history of the entire system. Moreover, spectroscopy of the tertiary can constrain its effective temperature, surface gravity, metallicity, mass, and evolutionary state, providing an independent stellar-evolution constraint on the age and history of the system. Combining these measurements with the exceptionally precise dynamical information available from pulsar timing should make it possible to test whether a single evolutionary scenario can simultaneously account for the MSP--helium-WD binary, the surviving tertiary, and the present three-dimensional architecture of this remarkable HT (\citealt{PCZ2012, Freire2026}; \citetalias{YangHT}).

Our SA formulation, together with the multi-technique follow-up strategy advocated here, provides a natural framework for addressing several of these outstanding questions concerning the formation and evolution of the system, while simultaneously exploiting PSR~J0435$+$3233 as a laboratory for precision tests of gravity. As a critical first step, the singly averaged \texttt{ELL1H} timing model presented in this work is currently being implemented in the pulsar-timing software package \texttt{PINT}~\citep{PINT1, PINT2}. By combining the dynamical information from high-precision FAST timing with independent spectroscopic and astrometric constraints on the tertiary, it should become possible to reconstruct the evolutionary history and three-dimensional architecture of the system while searching for subtle departures from general-relativistic dynamics, including possible violations of the universality of free fall. Detailed investigations along several of these directions are currently underway.

\begin{acknowledgments}
We thank Gerhard Sch\"afer for many helpful discussions. A.G., V.S., and S.B. acknowledge the support of the Department of Atomic Energy, Government of India, under Project ID No. RTI 4002. A.G. gratefully acknowledges support from the CAS President's International Fellowship Initiative (2026PVA0020). D.L. acknowledges support from the National Natural Science Foundation of China (grant nos. 12288102. P.W. acknowledges support from the National Natural Science Foundation of China under grant no. U2031l17, the Youth Innovation Promotion Association, CAS (id. 2021055). N.W. is supported by the National Natural Science Foundation of China (grant Nos. 12288102) and  the Major Science and Technology Program of Xinjiang Uygur Autonomous Region (grant nos. 2022A03013).  J.Y. is supported by the National Key Research and Development Program of China (No. 2022YFC2205203), and the Tianshan elites programme (grant no. 2023TSYCTD0013). D.Z. is supported by the Xinjiang Tianchi Talents Program (Young Doctor Program). Q.W. supported by the Youth Guidance Project of Zunyi Normal University (Grant No. QN(2026)09-(01)).

During the preparation of this manuscript, we used ChatGPT, DeepSeek, and NotebookLM for data analysis and visualization suggestions, preliminary plots, and minor grammatical and stylistic editing. All scientific analysis, interpretation, and conclusions were independently verified by the authors.
\end{acknowledgments}

\bibliography{references}{}
\bibliographystyle{aasjournalv7}

\end{document}